\documentclass[fleqn,usenatbib]{mnras}

\usepackage{graphicx}
\usepackage{color}
\usepackage{multirow}
\usepackage{amssymb,amsmath}
\usepackage{indentfirst}
\usepackage{leftidx}
\usepackage{CJK}

\makeatother

\newcommand{\kms}{\rm ~km~s^{-1}}

\newcommand{\Ha}{H$\alpha$}
\newcommand{\Pb}{Pa~$\beta$}

\title[Electron scattering in interacting supernovae]{Electron scattering wings on lines in interacting supernovae}
\author[C. Huang \& R. Chevalier]{
  Chenliang Huang (黄辰亮), $^{1,2}$ \thanks{E-mail: ch4de@virginia.edu}
  Roger A. Chevalier, $^{1}$
  \\
  $^{1}$ Department of Astronomy, University of Virginia, P.O. Box 400325, Charlottesville, VA 22904-4325, USA \\
  $^{2}$ Dept. of Physics and Astronomy, University of Nevada, Las Vegas, P.O. Box 454002, Las Vegas, NV 89154-4002, USA
}

\pubyear{2017}

\begin{document}
\begin{CJK*}{UTF8}{gkai}
\title[Electron scattering in interacting supernovae]{Electron scattering wings on lines in interacting supernovae}
\author[C. Huang \& R. Chevalier]{
  Chenliang Huang (黄辰亮), $^{1,2}$ \thanks{E-mail: ch4de@virginia.edu}
  Roger A. Chevalier, $^{1}$
  \\
  $^{1}$ Department of Astronomy, University of Virginia, P.O. Box 400325, Charlottesville, VA 22904-4325, USA \\
  $^{2}$ Dept. of Physics and Astronomy, University of Nevada, Las Vegas, P.O. Box 454002, Las Vegas, NV 89154-4002, USA
}

\label{firstpage}
\pagerange{\pageref{firstpage}--\pageref{lastpage}}
\maketitle

\begin{abstract}
We consider the effect of electron scattering on lines emitted as a result of supernova interaction with a circumstellar medium, assuming that the scattering occurs in ionized gas in the preshock circumstellar medium.  The single scattering case gives the broad component in the limit of low optical depth, showing a velocity full width half maximum that is close to the thermal velocities of electrons.  The line shape is approximately exponential at low velocities and steepens at higher velocities.  At higher optical depths, the line profile remains exponential at low velocities, but wings strengthen with increasing optical depth.  In addition to the line width, the ratio of narrow to broad (scattered) line strength is a possible diagnostic of the gas.  The results depend on the density profile of the circumstellar gas, especially if the scattering and photon creation occur in different regions.  
We apply the scattering model to a number of supernovae, including Type IIn and Type Ia-CSM events.
The asymmetry to the red found in some cases can be explained by scattering in a fast wind region which is indicated by observations.

\end{abstract}

\begin{keywords}
circumstellar matter --- shock waves --- supernovae: general
\end{keywords}

\section{INTRODUCTION}

The effects of electron scattering on emission lines have been discussed in various contexts.
One is an explanation for the broad emission lines observed in Seyfert galaxies
\citep{weymann70,laor06}, although this is not currently the preferred explanation for
broad lines.
In an expanding medium, electron scattering is expected to produce
a wing on the red side of an emission line.
\cite{auer72} noted the possible relevance of this process to Wolf-Rayet stars and
Seyfert galaxies 
\cite[see also][]{hillier91}.
In the context of supernovae, \cite{fransson89} examined the effect of electron scattering on lines
formed in the freely expanding ejecta during the nebular phase.
As above, scattering in the radially expanding gas gives a red wing to the line.
In the case where the thermal velocities of electrons dominate, the scattering primarily has a 
symmetric broadening effect about zero velocity.
This is the case studied by \cite{chugai01} for application to early spectra of the
Type IIn supernova SN 1998S.
In this scenario, after the supernova shock wave has broken out of the progenitor star and ionizing
radiation from the shock region is able to ionize the surroundings, the circumstellar 
medium around the supernova shock has substantial optical depth to electron
scattering.
This situation can occur in a supernova with a dense circumstellar medium because a
viscous shock is expected to form when the optical depth to the shock is
$\tau \sim c/v_s$, where $c$ is the speed of light and $v_s$ is the shock velocity
\citep[e.g.,][]{chevalier11,katz11}.
A $10,000\kms$ shock wave breaks out at $\tau = 30$.

The observational signature of electron scattering is broad wings (1000's of km s$^{-1}$) on 
a narrow line feature with velocities
that are characteristic of the circumstellar medium.
Electron scattering line profiles have been calculated and applied to a number
of observed supernovae, including SN 1998S \citep{chugai01}, SN 2005gj \citep{aldering06},
SN 2011ht \citep{humphreys12}, and SN 2010jl \citep{fransson13,borish13,dessart15}.
In addition, electron scattering has been mentioned as probably important for
other supernovae, such as SN 2008am \citep{chatzopoulos11}.
Supernovae with narrow spectral lines that have electron scattering wings are naturally classified
as Type IIn (narrow line).
However, electron scattering is probably not a factor in all Type IIn supernovae because
they have a range of circumstellar densities, and
at late times the electron scattering optical depth is expected to become small as
the supernova shock wave sweeps up the scattering circumstellar gas.
The mass motions then become the dominant factor in the line profiles.

Our primary aim here is to treat the line wings outside of the unscattered line emission.
We assume that electron scattering is the only opacity and neglect line opacity, as did \cite{auer72} and \cite{chugai01}, recognizing that there may be significant optical depth in the line, especially for \Ha.
Calculations including these effects were carried out for Wolf-Rayet stars \citep{hillier91}, 
SN 1994W \citep{dessart09,dessart16} and SN 2010jl \citep{dessart15}.
This calculation requires the treatment of the non-equilibrium level populations in the radiation field of the object.
There is considerable uncertainty in these quantities for supernovae.
The physics of the scattered line wings is relatively straightforward and is the case studied here.
The aim is to find diagnostics provided by observations of the broad line component that are relatively model independent.

Although electron scattering has frequently been invoked for broad lines in
interacting supernovae, there has not been a systematic investigation of the
line properties.
We investigate here the dependence of the line profiles on various
parameters, including the optical depth, the density distribution of the circumstellar
gas, and the velocity profile of the circumstellar gas.
These properties provide potential diagnostics of the supernova interaction.
The basic theory and results are presented in Section 2.  
Applications to observed supernovae are in Section 3.
The results are discussed in Section 4.

\section{SCATTERING IN A CIRCUMSTELLAR MEDIUM}

\begin{figure}
\includegraphics[width=\columnwidth]{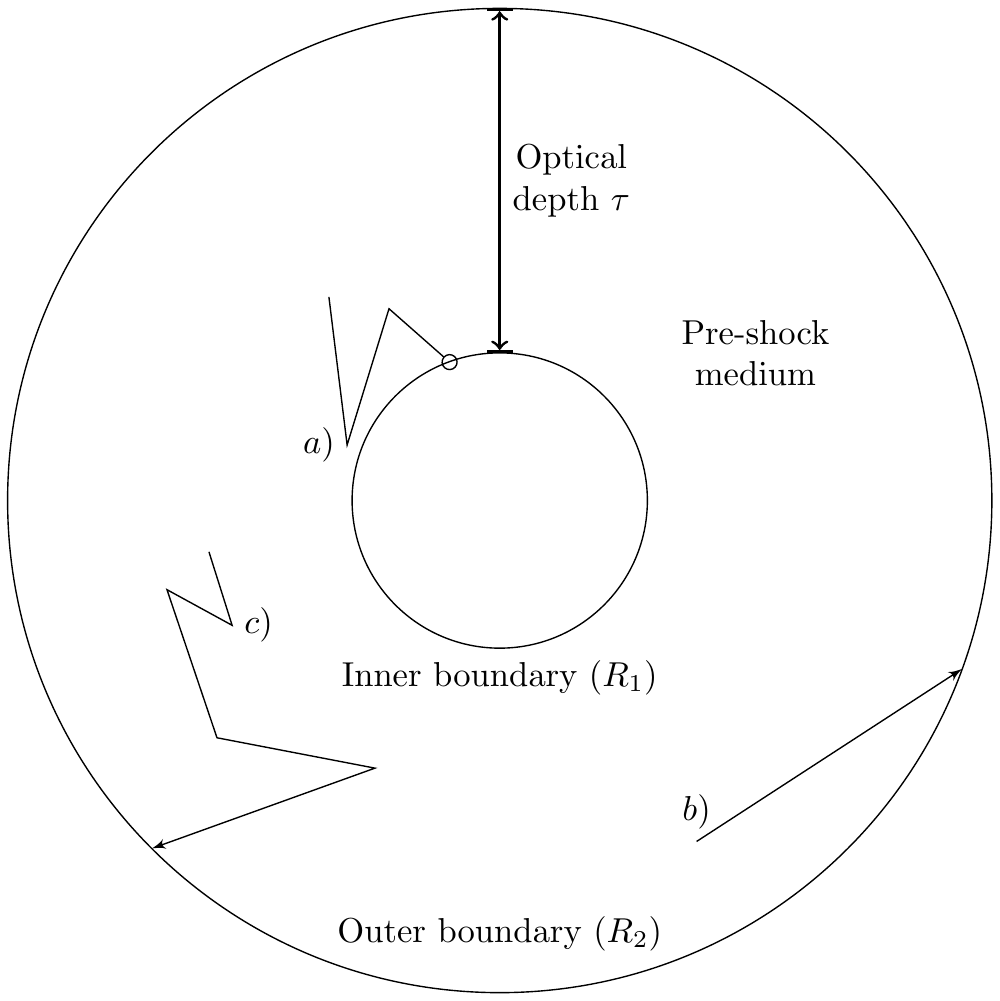}
\caption{Illustration of the scattering region, which extends from $R_1$ (inner boundary) to $R_2$ (outer).
A photon generated in the wind gas can \emph{a)} be absorbed by the inner boundary, \emph{b)} escape the medium from the outer boundary without any scattering,  contributing to the narrow component of the spectrum, and \emph{c)} escape from the outer boundary after scattering a certain number of times,  contributing to the broad component. 
}
\label{sketch}
\end{figure}

As in many previous treatments of electron scattering, we used a Monte Carlo scheme
to calculate the effects of scattering.
Because our primary application is to the optical spectra of Type IIn supernovae, we
 assume scattering in the Thomson limit, using the Thomson scattering differential cross section relation for each photon scattering.
We assume the scattering region is composed of fully ionized hydrogen with constant temperature T.
The effect of polarization is ignored; our models are spherically symmetric.
To make the simulation more efficient, statistical weights were assigned to the photons
as described in section 9.3 of \cite{pozdnyakov83}.
The simulations were carried out with the scattering medium
between an inner radius $R_1$ and an outer radius $R_2$ (Fig.\ \ref{sketch}).
The electron scattering optical depth $\tau$ through the medium is along the radial direction.
A test of the code was provided by the analytical solution of \cite{weymann70};  we
found good agreement of the Monte Carlo calculations with this solution.

\subsection{Single scattering limit}

\begin{figure}
\includegraphics[width=\columnwidth]{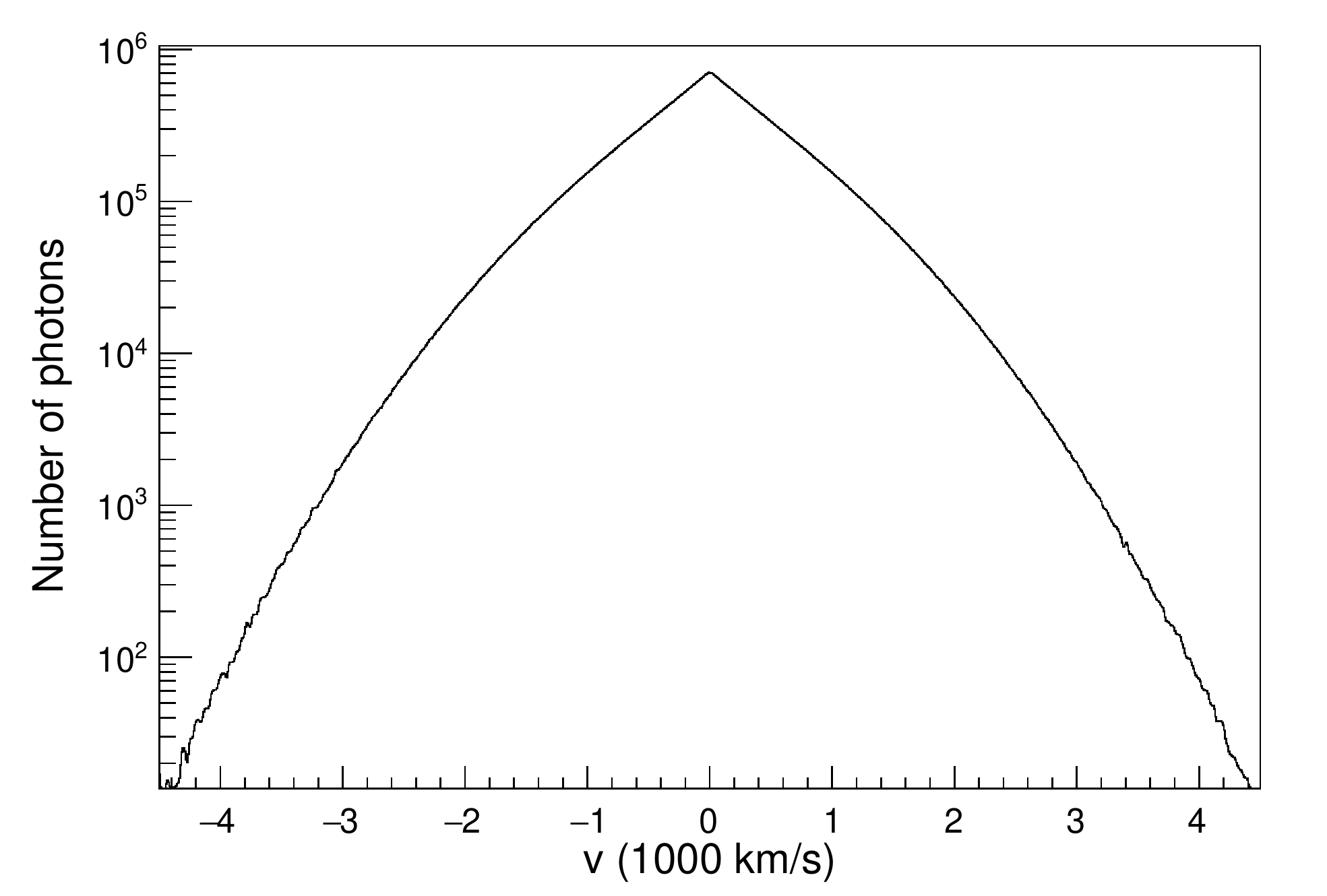}
\caption{Line profile for single scattering by thermal electrons with $T=20,000$ K.
}
\label{single}
\end{figure}

As an initial case, we calculated the situation where there is only single scattering
by a thermal distribution of electrons.
This case depends only on the broadening by thermal electrons and does not depend
on the parameters for the circumstellar gas other than the temperature.
In a realistic situation, the broad line profile should approach this case in the
low optical depth limit.
Fig.\ \ref{single} shows the resulting distribution of scattered photons for an assumed electron
gas temperature of 20,000 K.
It can be seen that the profile over the first factor $\sim 4$ in flux is approximately
exponential and steepens from an exponential beyond that.
The full width at half maximum (FWHM) of the line can be expressed as $\Delta v=913 (T/20,000{\rm~K})^{1/2}\kms$.
 To obtain a robust measurement of the FWHM, we fit an exponential to a $\sim 100 \kms$ region near the profile peak to determine the maximum flux, and an exponential near the half maximum bin to determine the width.  
 This method helped to account for fluctuations in the Monte Carlo results.
For comparison, the mean thermal velocity of an electron at 20,000 K is $954\kms$.
The line profile compares well with the single scattering result shown in Fig.\ 2 of
\cite{sunyaev80},
whose result has some asymmetry because 5.1 keV photons are considered and the
situation is not fully in the Thomson limit.
 In the following discussion, we bin non-scattered and scattered photons (narrow and broad component) separately in order to measure a well defined FWHM of the broad component.

\subsection{Stationary circumstellar medium} \label{sec_basic}

Our aim is to present physically plausible situations where electron scattering plays
a role.
We begin by treating simpler situations and proceed to more complex ones.
The initial case is an ionized wind with density profile $\rho_w\propto r^{-2}$,
as expected for a steady wind.
The wind velocity is assumed to be negligible so the circumstellar gas is
effectively stationary.
The inner boundary is assumed to be an absorbing sphere; this might be a 
shocked shell.
Most of the contribution to the optical depth and emissivity comes from
close to the inner boundary for this density profile.
We assume that the wind gas is responsible for both the emission of line photons
and the subsequent scattering of these photons.
The line emissivity is taken to be $\propto n^2$, where $n$ is the gas density.
Under these conditions, the broad component of the line profile depends only
on the optical depth to the absorbing sphere, $\tau$, and the radial extent of the
scattering electrons.

\begin{figure}
\includegraphics[width=\columnwidth]{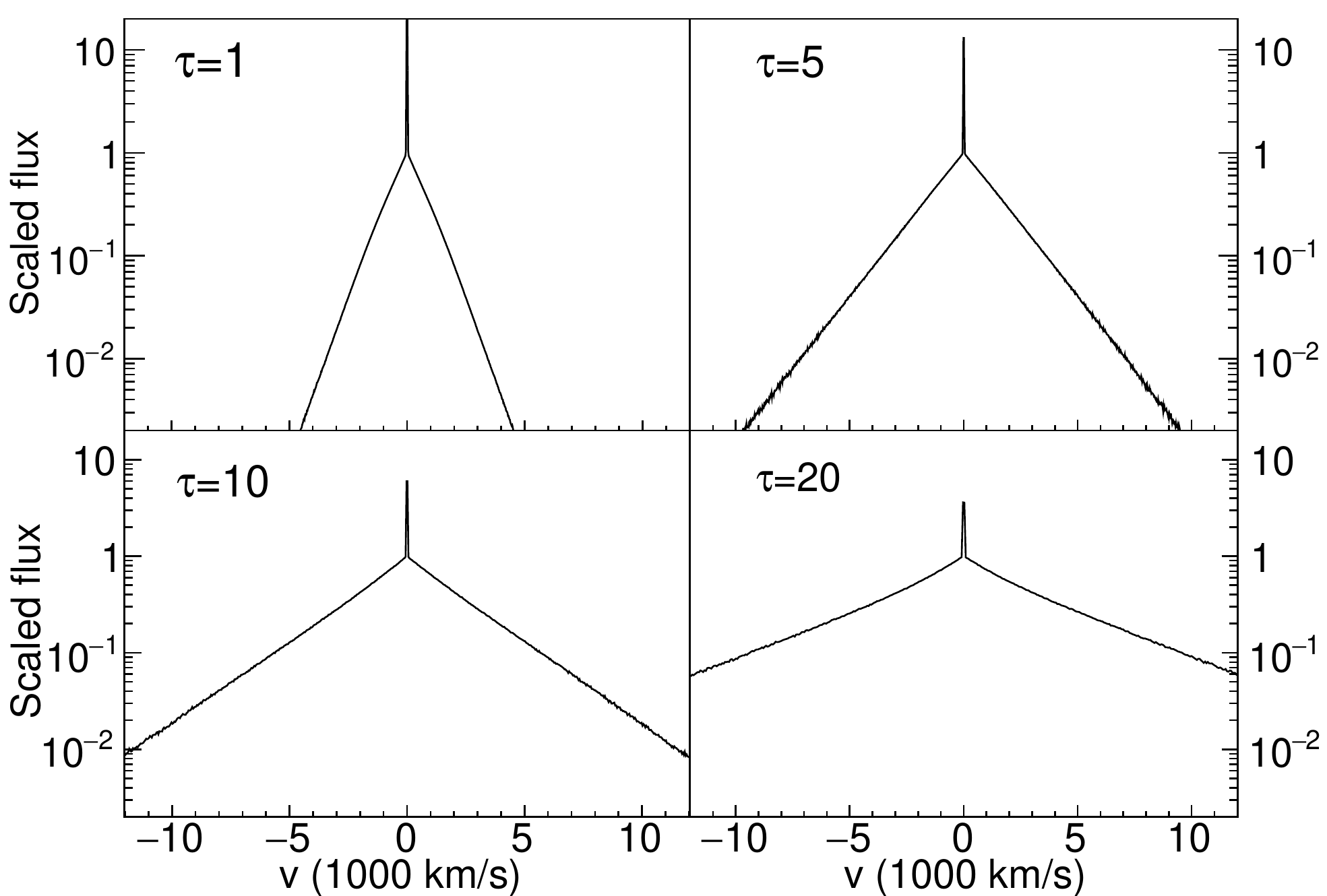}
\caption{Line profile results plotted on a log scale, for a circumstellar medium with $T=20,000$ K, $n\propto r^{-2}$, $R_2/R_1=10^3$, and a static medium. 
The optical depths $\tau=1,5,10,20$ are shown. 
}\label{basic}
\end{figure}

\begin{figure}
\includegraphics[width=\columnwidth]{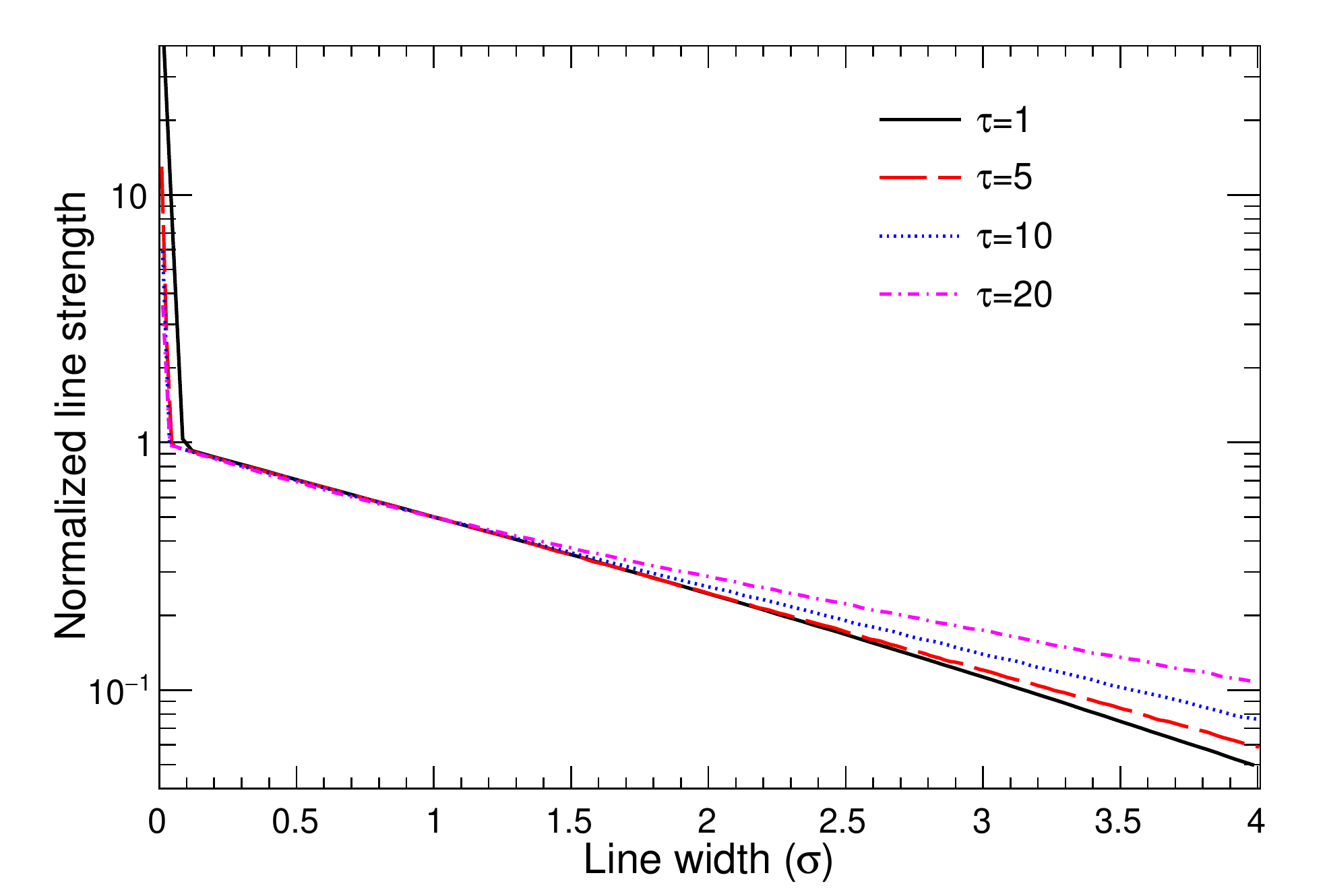}
\caption{Normalized line shape, with the lines  folded about 0 velocity. An  $n\propto r^{-2}$ density profile is assumed and $R_2/R_1=10^3$.  
The $x$-axis, $\sigma$, is the line half width in  units of the half width half maximum, and the $y$-axis is normalized to the peak of the broad component.
}
\label{normalize}
\end{figure}

\begin{figure}
\includegraphics[width=\columnwidth]{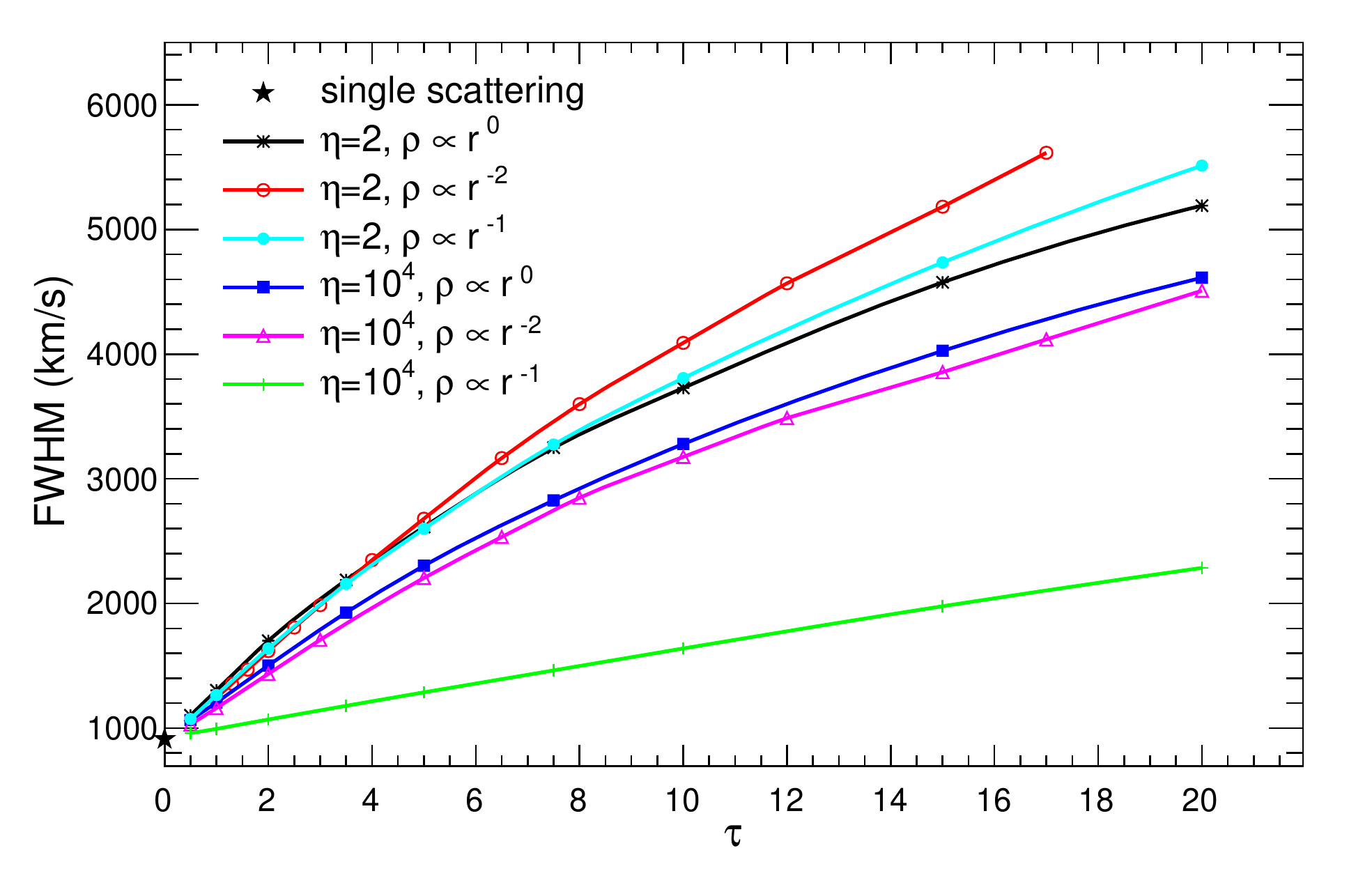}
\caption{Relation of FWHM  vs. $\tau$, assuming  $T=20,000$K, $n\propto r^{-2}, r^{-1}, r^0$, and two different radial extents $\eta=R_2/R_1$.  For comparison, the FWHM of the single scattering case, $913\kms$,  is shown as a black star.
}
\label{FWHM_more}
\end{figure}

Fig.\ \ref{basic} shows the resulting line profiles for different values of $\tau$.
The ratio $R_2/R_1$ is $10^3$.
As expected, higher optical depths lead to multiple scattering and a broader broad
component than is obtained in the single scattering case.
There is also a narrow component that is made up of line photons that escape without scattering.
To characterize the width of the approximately exponential scattering wing, the FWHM of the broad (scattered) component is used here.
Fig.\ \ref{normalize} shows the normalized line shapes such that the line profiles have the same FWHM.
It can be seen that the line profile shapes are similar over the top part of the profile, 
but that
the line wings are relatively stronger at high $\tau$.
At low $\tau$, the line profile drops more rapidly than an exponential out in
the wings, while at high $\tau$, the profile drops less rapidly than an
exponential.
At $\tau\approx 10$, the profile remains exponential far out in the wings.
The results are shown for a particular temperature $T$, but the electron thermal
velocity is $\propto T^{1/2}$, so that the FWHM $\propto T^{1/2}$.
The FWHM of the broad component is shown in Fig.\ \ref{FWHM_more} as a function of $\tau$.
At low optical depth, the FWHM approaches the single scattering value.

\begin{figure}
\includegraphics[width=\columnwidth]{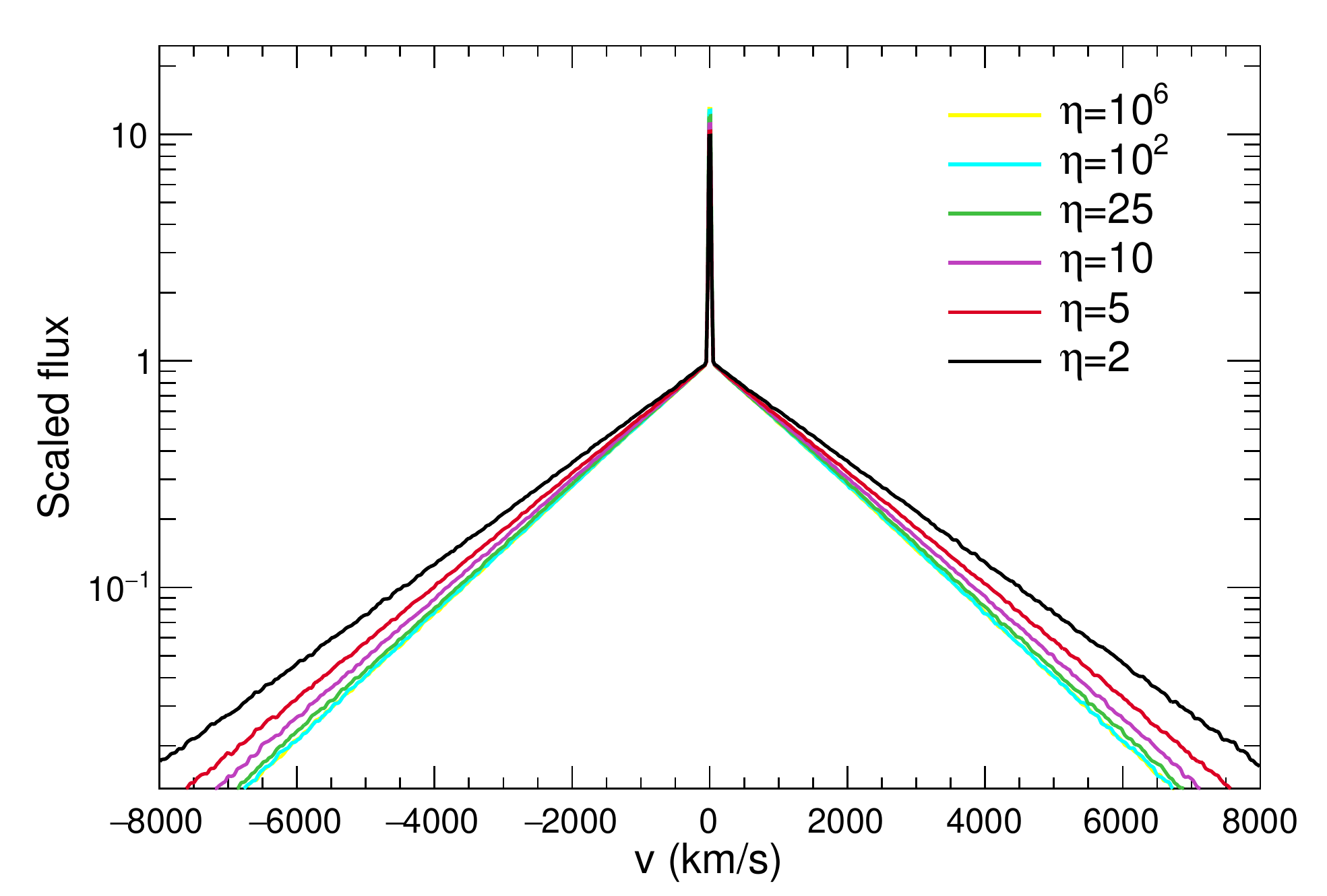}
\caption{The line profile for a series of outer to inner radius ratios ($\eta=R_2/R_1$) assuming an $n\propto r^{-2}$ medium.  The profile for $\eta=10^6$ overlaps with the profile for $\eta=10^2$.
}
\label{ratio_r2}
\end{figure}

One parameter is the radial extent of scattering electrons.
Fig.\ \ref{ratio_r2} shows the line profiles resulting from a number of different extents.
The profile is expected to depend only on the ratio $\eta =R_2/R_1$.
It can be seen that, provided that $R_2/R_1\ga 20$, the line profile depends weakly on the radius ratio.
This conclusion also holds for other power law density profiles except a negative power law index $ \sim -1$ (see below
for more discussion).
Fig.\ \ref{ratio_r2} also shows that the line formed over a narrow radial region, which approximates
a planar scattering layer, gives a broader line than in the case of a large radial extent.
The result can be understood in that a photon escaping from the vicinity of a spherical region inner boundary sees a slowly risng optical depth away from
a radial line while, in the planar case, the rise is more rapid.
The planar case thus leads to a higher mean optical depth and broader lines.

\begin{figure}
\includegraphics[width=\columnwidth]{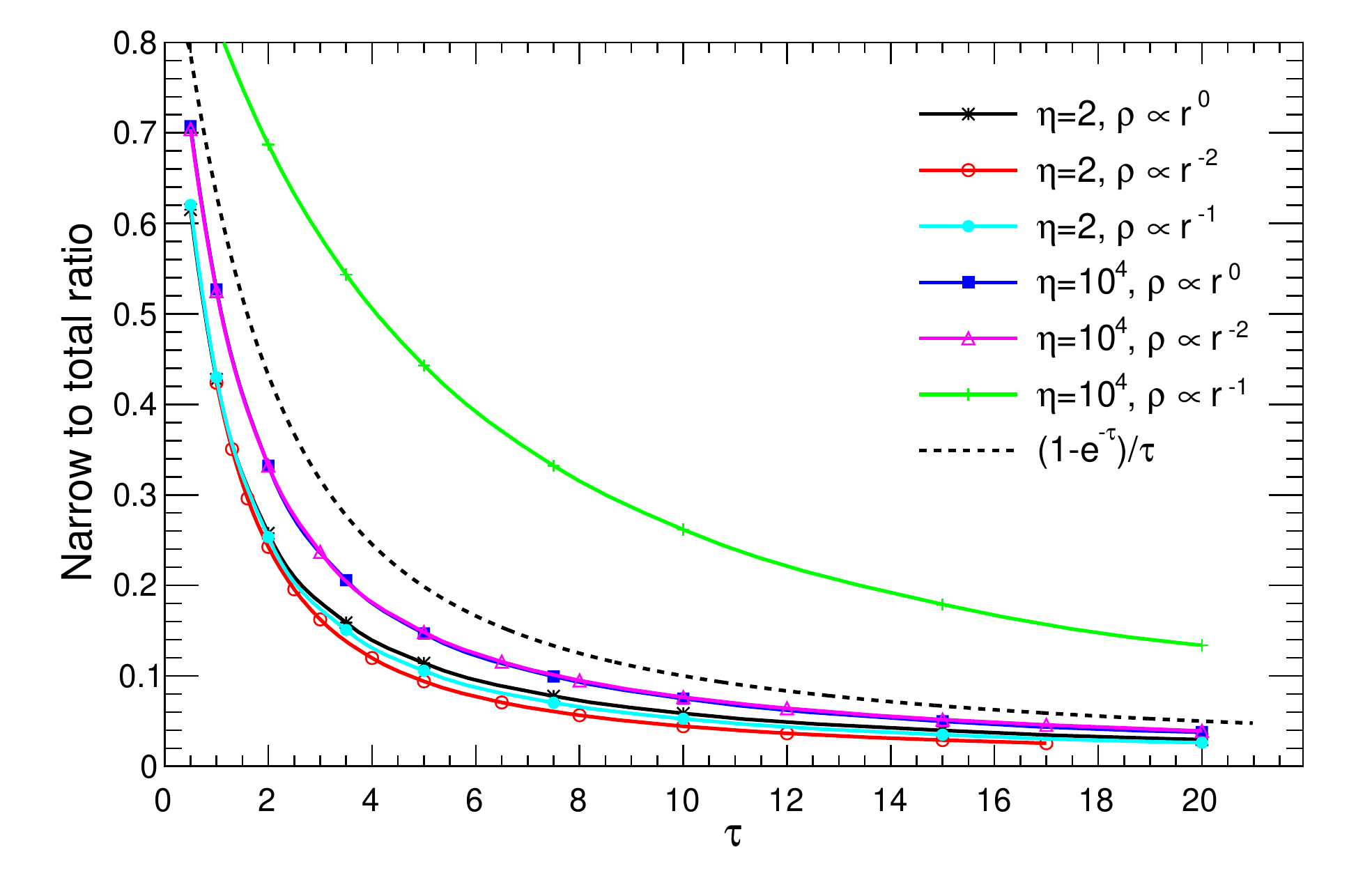}
\caption{Relation of narrow to total line ratio  vs. $\tau$, assuming  $T=20,000$K, $n\propto r^{-2}, r^{-1}, r^0$, and two different radial extents ($\eta=R_2/R_1$).   The dashed line shows the relation given by a simple analytic formula (see text).
}
\label{n2t_more}
\end{figure}

Fig.\ \ref{basic} shows the presence of the narrow component mentioned above.
The ratio of narrow to broad component depends on the escape probability
of the photons and is shown in Fig.\ \ref{n2t_more}.
We also give the approximate expression for the escape probability $[1-\exp(-\tau)]/\tau$;
it can be seen that the accurate value is somewhat below this estimate.
The approximate expression is only accurate under the assumption that the
photons are confined to move in the radial direction.
An assumption made in these simulations is that there are no effects of line opacity.
In Type IIn supernovae, the narrow \Ha\ component frequently shows 
optical depth effects, i.e. a P Cygni line profile \citep[e.g.,][]{kiewe12}, so the present
considerations do not strictly apply.
However, the results for the narrow line are potentially applicable to higher level H lines and other lines
with lower optical depth.

\begin{figure}
\includegraphics[width=\columnwidth]{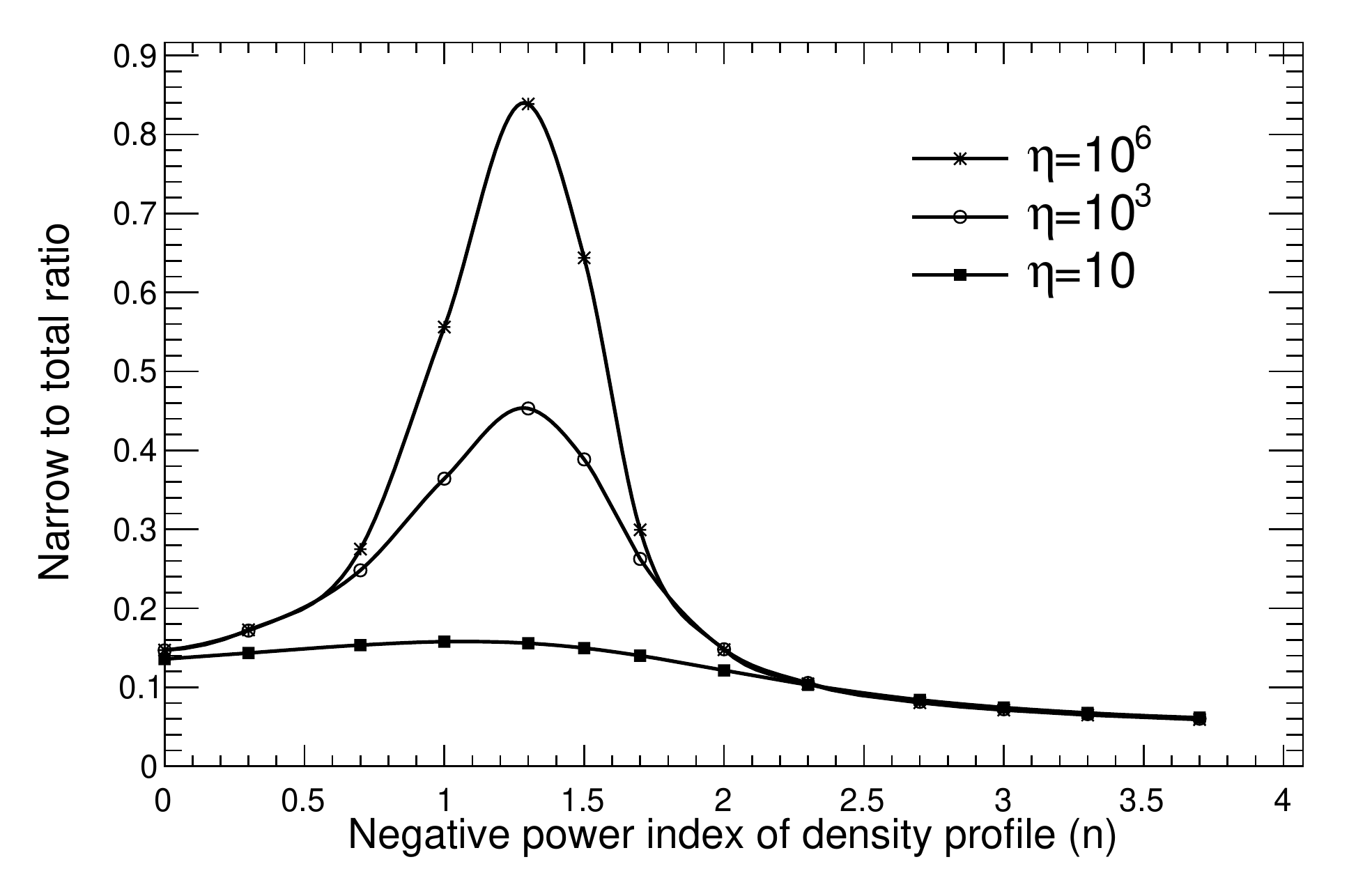}
\caption{Narrow to total line ratio vs. the power law of the density profile for 3 different sets
of radial ranges ($\eta=R_2/R_1$), assuming $\tau=5$ and $T=20,000$ K.  Here, $\eta$ is the ratio of outer radius
to inner radius for the scattering region.
}
\label{density_n2t}
\end{figure}

\begin{figure}
\includegraphics[width=\columnwidth]{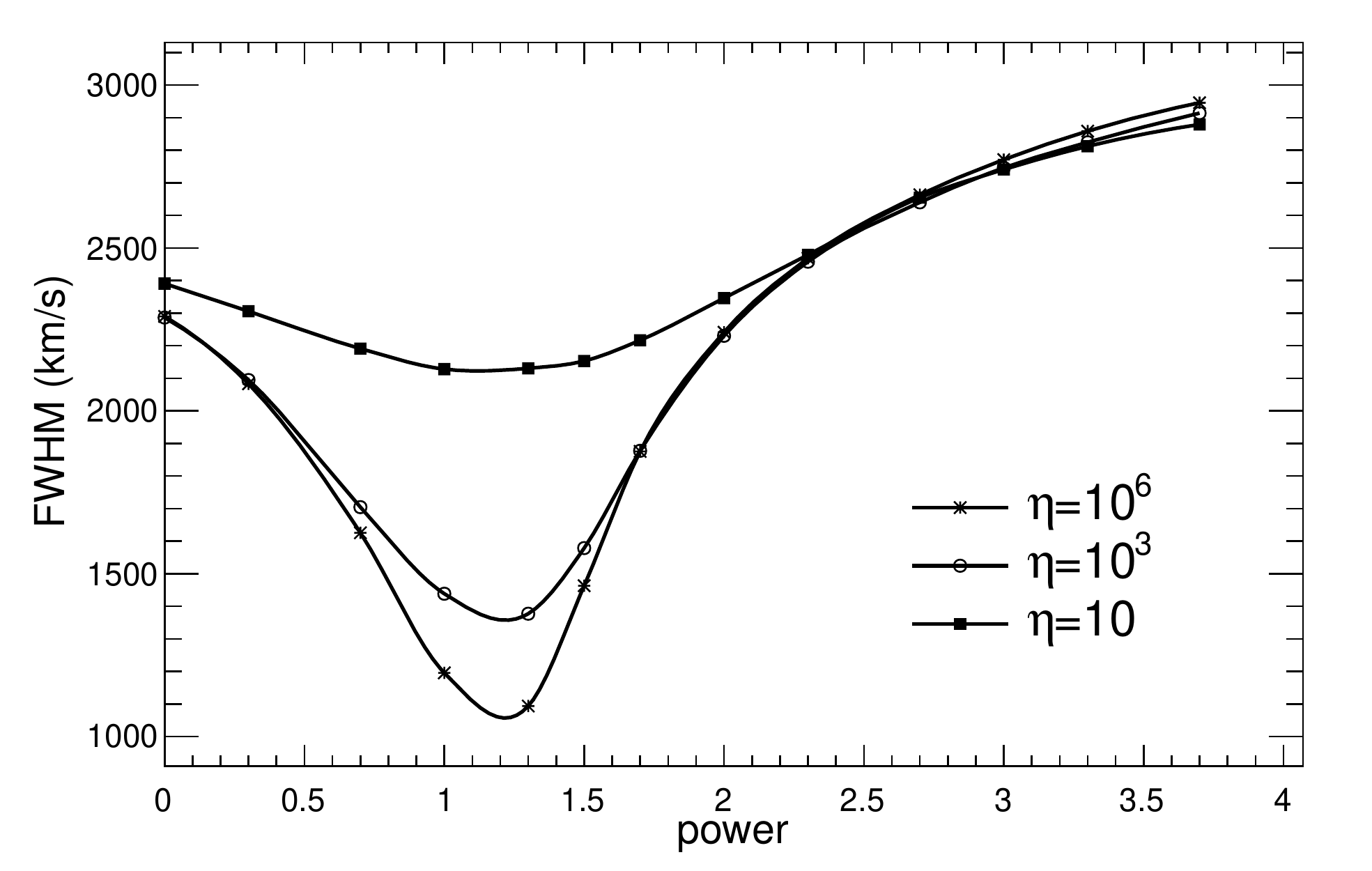}
\caption{Same as  Fig.\ \ref{density_n2t} but for the FWHM.
}
\label{density_FWHM}
\end{figure}

We have assumed a density profile that is appropriate to a steady wind from the
progenitor star.
However, the mass loss processes that give rise to the dense circumstellar medium
around a Type IIn supernova are not understood and are likely to be complex.
In order to check the sensitivity to the density profile, we undertook simulations
with a density profile $\rho\propto r^{-s}$, with $s$ in the range $0-3.7$ and $\tau =5$.
The FWHM and narrow to total line ratio are shown as a function of $s$ in Figs.\ \ref{density_n2t} 
and \ref{density_FWHM}.
It can be seen that, for the narrowest scattering layer, there is the least dependence on $s$.
In the limit that the layer is geometrically thin, the results are expected to be independent of $s$.
For $s=2$, already discussed, the insensitivity of the results to the radial extent
is clear; most of the contribution to scattering comes from layers that are close to $R_1$.
The same is true for $s>2$.
On the other hand, for a radially extended region with $s=0$, most of the scattering
occurs in layers close to $R_2$.
The results do not depend significantly on the position of $R_1$ provided that the region
is radially extended.
In comparing line profiles for the $s=0$ case and the $s=2$ steady wind case,
we find that the FWHM approximately agrees for the two cases, but that the outer
wings of the line profiles are elevated in the constant density case.

It can be seen in Figs.\ \ref{density_n2t} 
and \ref{density_FWHM} that there are significant deviations from the standard results
between $s=0$ and 2, which we interpret as follows.
The production of photons in a given logarithmic radial range is $\dot N\propto 4\pi R^2 \Delta R \epsilon$,
where $\Delta R\propto R$ and $\epsilon\propto n^2$ is the gas emissivity, so that
$\dot N\propto R^{3-2s}$.
Thus for $s<1.5$, most of the photons are produced at large radii, while for $s>1.5$, they are
produced at small radii.
For the optical depth $\tau \sim \int \rho\kappa dr$, $s=1$ is a critical value above which most
of the optical depth is contributed at small radii and below which most is contributed at
large radii.
The result is that for $s> 1.5$ both the photon production and the optical depth effects occur
at small radii, while for $s<1$, they both occur at large radii.
In the range $1<s<1.5$, most of the photons are produced at large radii, but the electron scattering optical depth
occurs at small radii.
The result is that the photons can escape more easily without scattering, as is shown
in Fig.\ \ref{density_n2t}.
At the same time, the rapid escape of photons leads to a narrower FWHM (Fig.\ \ref{density_FWHM}).

\subsection{Presupernova mass loss velocity}
\label{sec:wind}

\begin{figure}
\includegraphics[width=\columnwidth]{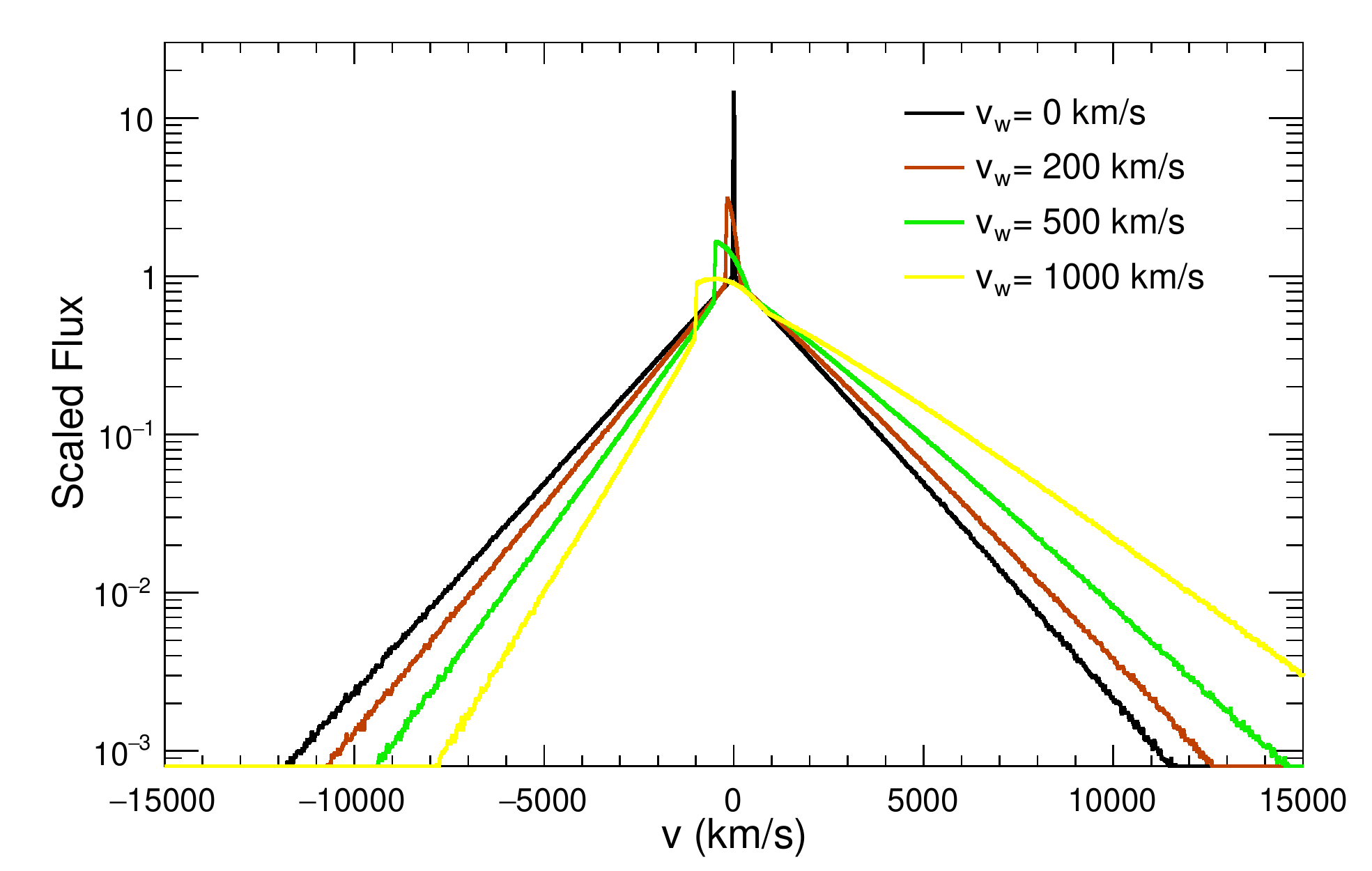}
\caption{Line profiles for $n\propto r^{-2}$, $\tau=5$, $R_2/R_1=10$, $T=20,000$, and various values of the wind velocity $v_w$, with constant outflow velocity.  
}
\label{same_vmax_r2}
\end{figure}

Up to this point, we have assumed that the scattering medium is stationary so that
the line broadening is entirely due to the thermal velocities of electrons.
In the actual case, the mass loss gas is expected to have some outflow velocity, $v_w$,
from the parent star, giving rise to an asymmetry to the red in the line profile \citep{auer72}.
In Fig.\  \ref{same_vmax_r2}, we show the line profiles for various values of $v_w$ for the standard case
with $\tau=5$, $s=2$, $R_2/R_1=10$, and $T=20,000$ K.
It can be seen that even for $v_w=200\kms$ there is a some asymmetry
introduced into the broad line profile, even though the thermal velocity of the 
electrons is $954\kms$.
We attribute the effect to the fact that the wind velocity gives a systematic redshift to
the scattering photons, while the thermal velocities are equally positive and negative
resulting in a diffusion in frequency.
The systematic redshift is due to the spherical divergence of the flow.
In the radial direction, the velocity is constant, so there is no systematic change in
photon energy.
For scattering in a nonradial direction, there is a velocity component away from
the photon source that gives a systematic redshift.
A similar situation occurs for electron scattering in the winds of Wolf-Rayet stars,
although in this case there may be a region where the wind accelerates in the radial
direction to its final velocity \citep{hillier91}; the acceleration is expected to
give rise to a systematic redshift of scattering photons.
As can be seen in fig.\ 3 of \cite{hillier91}, the red wing of the scattered component
is considerably stronger than the blue wing.

In the present case, there is no velocity gradient in the radial direction, so that
scattering away from a radial line is important for causing an asymmetry.
In addition, scattering at a great distance from the photon source gives rise to
a greater velocity difference and redshift.
Thus the case of $n\propto r^{-2}$ in an extended scattering region is especially
favorable for producing an asymmetry.
Photons are emitted at a deeper optical depth on average for a thicker scattering region, which could also contribute to the asymmetric profile.
If the scattering layer is narrow, the photons do not move far in angle before
escaping and thus the asymmetry is small.
For the case studied by \cite{chugai01}, $R_1=4\times 10^{14} $ cm and $R_2=9\times 10^{14} $ cm, so the narrow scattering region gives rise to a negligible asymmetry;  the wind velocity was taken to be $40\kms$.  
The asymmetry of the narrow peak weakens with increasing thickness of the scattering layer because fewer non-scattered (and redshifted) photons are absorbed by the inner boundary.

We considered the effect of the density power law index for the scattering gas ($\rho \propto r^{-2},~r^{-1},~r^0$).
For the broad component, the degree of asymmetry is comparable for the 3 density profiles.
The far blue wing becomes stronger for a flatter density profile.
For the uniform density case, most photons are created near the outer boundary.
Because of the density profile, even photons that escape horizontally must penetrate a large optical depth, so there are fewer narrow line
photons at zero and positive velocity.

We investigated variations in the optical depth.
If the medium is spatially thick, the asymmetry is insensitive to the optical depth.
However, if the medium is spatially thin, for a larger optical depth the asymmetry is slightly smaller because the photons cannot travel far in angle between two scatterings.
Because a thinner medium leads to a smaller asymmetry for the same expansion velocity, a larger wind velocity is required in fitting the same spectrum.  
As a result, the profile of a spatially thinner medium shifts to the blue more than a thicker medium which has the same asymmetry.
The sharp jump at the blue edge of the narrow component would be smeared out by the Gaussian convolution in an actual observed profile.

We also considered the line profiles in the linear velocity profile case where $v_w=v_b  (r/R_2)$, as might occur in explosive mass loss.
Here, $v_{b}$ corresponds to the velocity at the outer boundary.
This explosive mass loss scenario has been suggested for SN 2006gy \citep{smith10a}.
The profile is similar to a thin layer constant wind velocity case with a smaller optical depth.  
This is because the profile is dominated by the medium with the largest velocity which is in a relatively thin layer near the outer boundary.
Except for the shape of the narrow component, the profile is insensitive to the spatial thickness of the layer.

If the scattering medium is stationary, the top of the broad component has a sharp tip, like the shape shown in the single scattering profile (Fig.~\ref{single}). 
When scattered by an expanding medium, the broad component has a rounded top instead, shown as the green dotted line in Fig.~\ref{98s4}.
The rounded top becomes wider and lower as the wind velocity increases.
Because the width of the rounded top is always narrower than the narrow component, it does not show up in the overall spectrum.

\subsection{Radiative acceleration of circumstellar gas}

When a shock wave breaks out of the progenitor star, the radiation dominated shock
transition becomes broad as the radiation is able to diffuse away from the star.
The strong radiation field is expected to accelerate the circumstellar medium with a velocity
profile $\propto r^{-2}$ because of the spherically diverging radiation flux.
\cite{chugai01} used a velocity profile
\begin{equation}
v(r)= v_w+v_{sh}\left(\frac{R_{1}}{r}\right)^2,
\label{v_profile}
\end{equation}
where $v_w$ is the velocity of the presupernova wind and $v_{sh}$ is the velocity from
acceleration at the inner boundary $R_1$.  

\begin{figure}
\includegraphics[width=\columnwidth]{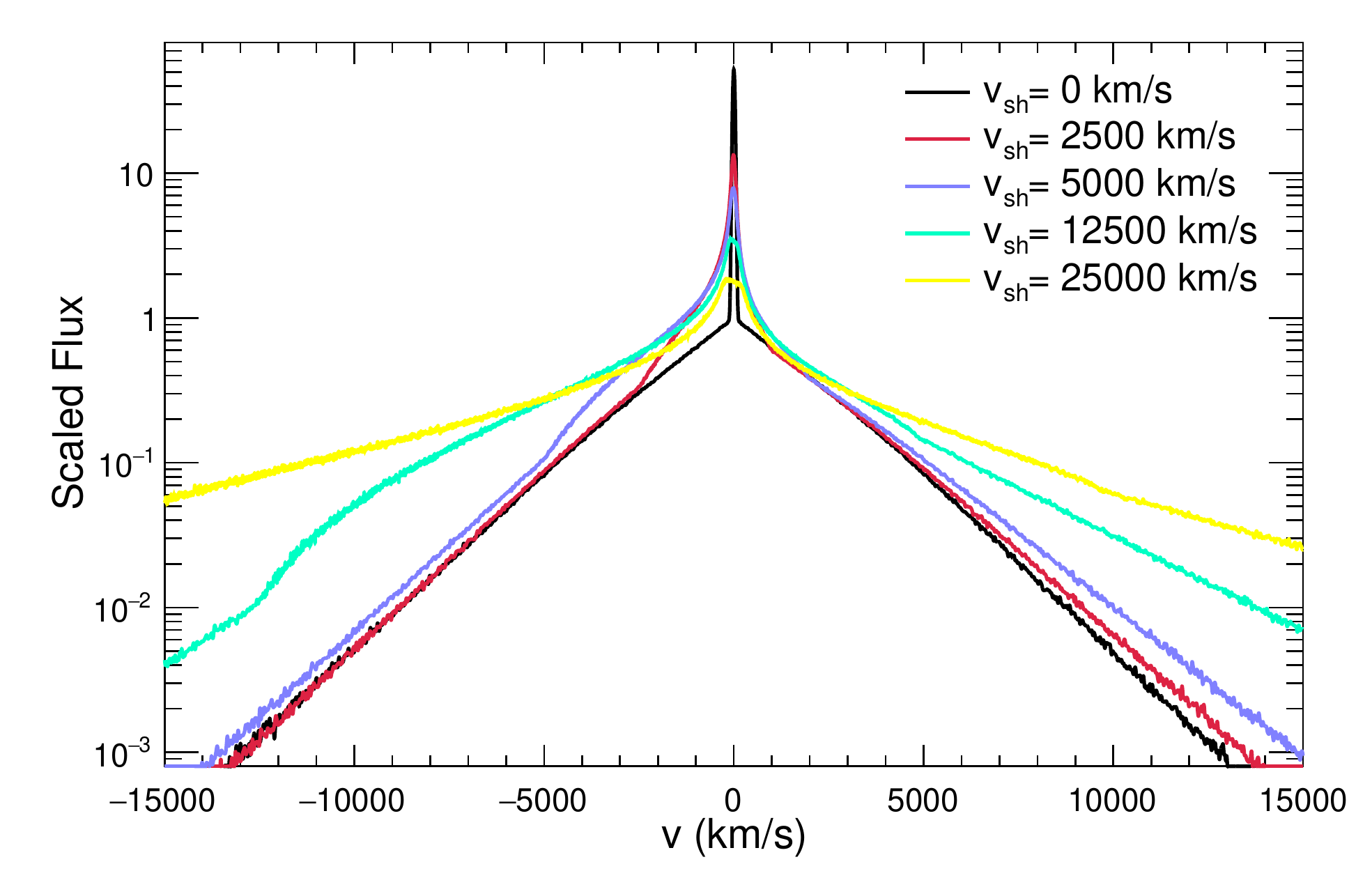}
\caption{Line profiles for $n\propto r^{-2}$, $\tau=1$, $R_2/R_1=10$, $T=125,000$ K, and various values of the preshock velocity $v_{sh}$ determined by radiative acceleration of the gas.
  The wind velocity $v_w=0$.
}
\label{tau1_same_vmin}
\end{figure}

Some insight into the effect of $v_{sh}$ can be obtained from examination of the $\tau =1$ case.
Fig.\ \ref{tau1_same_vmin} shows the profile with $\eta=10$ and $T=125,000$ K.
The reason for choosing this unphysically high temperature is to match its width with the larger optical depth case shown in Fig.~\ref{same_vmax_r2}.
The line profile is substantially affected out to $v_{sh}$, especially on the blue side because of occultation effects on the red side (see the $v_{sh}=5000\kms$ case).
An inflection appears at the connection point between the region dominated by the narrow component and by the broad component.
As noted by \cite{chugai01}, the effects become especially significant for $v_{sh}>1000\kms$.
In the velocity range $|v|<v_{sh}$, the main contribution is from unscattered photons and there is an asymmetry to the blue.
For $|v|>v_{sh}$, the flux is dominated by scattered photons, whose properties are primarily determined by the thermal velocities of electrons.
If the optical depth $\tau>2$, the narrow component smoothly transitions to the broad component and the inflection point disappears because the broad component takes over at a smaller $|v|$.

A larger $v_{sh}$ leads to a sharper drop on the red side of the narrow component and a flatter slope on the blue side, because most unscattered photons are emitted in a small optical depth region that is close to the outer boundary where the expansion velocity is small.
  This effect is less dramatic if the optical depth is small or the radius ratio is small.

For a geometrically thin layer, all the gas is fast moving, so the non-scattered component is broad.
For a thick layer, the outer part of the medium is almost static so the narrow component has a sharp peak (Fig.\ \ref{tau1_same_vmin}).
Especially when $\tau >1$, few photons near the inner boundary can escape without scattering.
The geometrical thickness has a small effect on the broad profile for a thick layer ($\eta\gtrsim 3$).
On the one hand, the outer part of the thick layer moves very slowly; on the other, a thick layer leads to
a stronger spherical divergence effect and more photons are emitted at a deeper optical depth.
These two effects basically cancel each other.

For a thick layer with uniform density or an $r^{-1}$ density profile, most photons are emitted in the outer
part of the medium, where the velocity is small if there is no wind velocity.
All the profiles are the same.

\subsection{Effects of continuum absorption}

\begin{figure}
\includegraphics[width=\columnwidth]{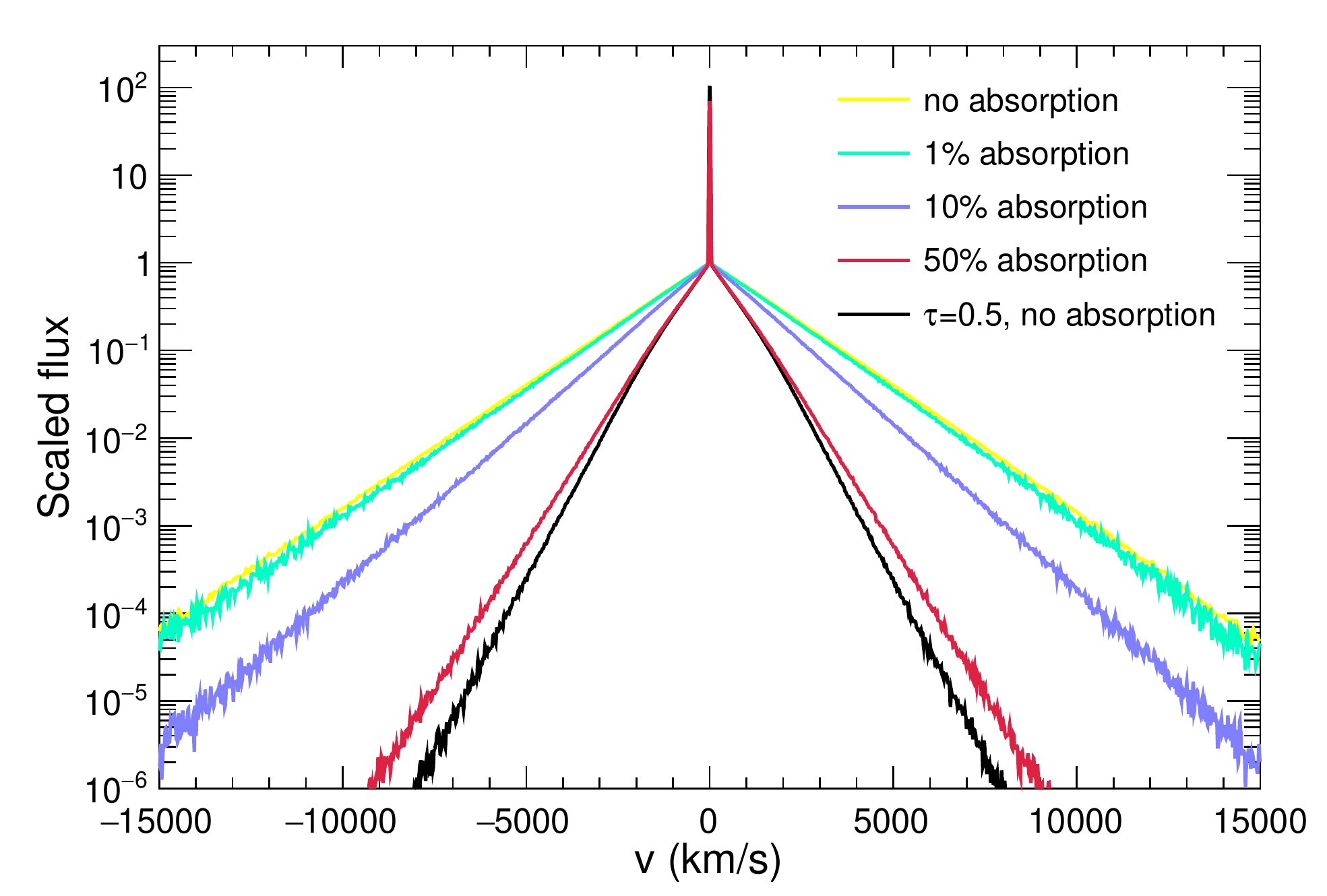}
\caption{The variation of the line profile with continuum absorption, assuming $\tau=5$, $T=20,000$ K,
$n\propto r^{-2}$, and a stationary circumstellar medium.  The line profile for $\tau=0.5$ is shown for comparison.
}
\label{abs_tau5}
\end{figure}

The calculations in the previous sections assume only scattering, with no absorption.
In order to estimate the effect of absorption, we introduced an absorption probability
parameter that is defined as the probability that a photon is absorbed for each
collision event.
Fig.\ \ref{abs_tau5} shows that the effect of continuum absorption is very similar to a smaller
optical depth with no absorption.
We note that continuum absorption does not give rise to a weakening of red emission,
as typically occurs in the supernova case because of absorption of emission from
the receding part of the explosion.
Here, the line broadening is due to the thermal velocities of the electrons and not
to the overall expansion.

In the absence of dust, the relevant absorption processes are free-free and bound-free
absorption.
As discussed by \cite{chugai01}, both of these are expected to be small for the
physical conditions of interest.

In principle, dust is also a possible source of continuum absorption and has been suggested
as the source of an asymmetric broad line in SN 2010jl \citep{smith12,gall14}.
As discussed above, this would require that the explosive motions play a role in the
line broadening.
In addition, at early times the radiation and gas temperatures remain $>5000$ K so that
the conditions are not appropriate for the formation of dust.
\cite{fransson13}  have discussed the issues with inferring the presence of dust in the supernova.

\section{COMPARISON WITH OBSERVATIONS}

The models calculated here were compared to observed supernova line profiles,  which were
downloaded from the WISeREP repository of supernova spectra \citep{yaron12}.
Spectra of Type IIn supernovae were chosen that had a well-observed H line which shows distinct broad and narrow components.
The supernova spectra that were used are listed in Table \ref{params} and are shown in Figs.~\ref{98s4} --  \ref{11ht}.       
The ages given in Table~\ref{params} are from the time of discovery.
We chose the H lines which have the best data available and a smooth continuum. 
For most cases, the \Ha\ line is studied.

One issue with modeling the \Ha\ line is that there can be interference on the red side of the line with the He I $\lambda$6678 line, which is at a wavelength corresponding to $\sim 5300\kms$ relative to \Ha.
The [\ion{N}{ii}] $\lambda$6611 and [\ion{N}{ii}] $\lambda$6482 emission lines may also present on the red and blue side of the \Ha\ respectively, corresponding to $\sim 2200\kms$ and $\sim 3700\kms$ relative to \Ha; the [\ion{N}{ii}] lines are probably from the host galaxy.
The references in Table \ref{params} are to the papers that presented the spectra we are using.
Where possible, we use the blackbody temperature estimates at those times to determine the background emission.

Another issue is that it commonly shows a narrow P Cygni line profile in the spectra of Type IIn SNe, showing that there are line optical depth effects in the narrow line.
Our models assume an optically thin medium and thus are primarily directed at the broad (scattered) line component.

The models were calculated as follows.  
The scattering occurs in the circumstellar medium outside of the supernova shock wave.
The gas is thus heated and ionized by the energetic radiation from the postshock region.
We considered a spherically symmetric isothermal circumstellar medium with a $r^{-2}$ density profile and a velocity profile described by Equation~(\ref{v_profile}).
We preferred a uniform expanding velocity, or $v_{sh}\approx 0$ in the fits, especially for the spectra at late time, because radiative acceleration is important around the time of shock breakout, which occurs at an early phase.
The \Ha\ photons are emitted by the same medium due to radiative recombination, so the emissivity is proportional to the density squared.  

The fitting parameters considered were gas temperature $T$, $\tau$, $\eta$, $v_w$, $v_{sh}$ and redshift $z$.
The calculated emission lines were convolved with Gaussian profiles with FWHM equal to the spectral resolution.
The preferred fitting parameters are listed in Table~\ref{fit_param}.
The listed FWHM values are the measured values for the broad components of the comparison models.
  For an expanding medium, the peak of the broad component is rounded, shown as the green dotted line in Fig.~\ref{98s4}.  
  As a result, the FWHM of the broad component is strongly model dependent.
  In order to reduce the sensitivity on the fitting model in characterizing the width of the broad component, and make the value applicable to be compared to the observational quantity, we measured an equivalent FWHM defined as following.
  As shown in Section~\ref{sec_basic}, the line wing of the broad component is approximately exponential.
  We fit the exponential $a e^{bv}$ to both line wings of the best fit model that has been convolved with the spectral resolution, where $v$ is the frequency to line centre in velocity units.
  Because the exponential index $b$ varies slowly with $v$, the exponential fit is performed near the half maximum bin of the broad component.
  However, if the model has a high wind velocity or low spectral resolution, e.g., SN 2012bq and SN 2008cg, the narrow component can extend to the half maximum bin of the broad component.
  For these two cases, we fit the exponential further in the line wing to avoid contamination from the narrow feature.
  With the measured exponential index on both sides and by assuming the broad component is made up of two exponential wings, we can define the equivalent FWHM:
  \begin{equation}
    \label{eq:FWHM}
    FWHM=\ln 2 (1/b_b - 1/b_r),
  \end{equation}
  where $b_l$ and $b_r$ stand for the fitted exponential indexes on the blue and red side respectively.
  Due to the selection of best fit model, as well as the exponential index measurement, the uncertainty in the FWHM measurement is about $100 \kms$.
  This equivalent FWHM of each observation is listed in Table~\ref{fit_param}.

The two main model parameters determining the width of the broad component are the electron scattering optical depth $\tau$ and the temperature $T$ of the gas.
The line profile by itself cannot be used to determine $\tau$ because sensitivity to $\tau$ generally occurs far out in the wings where the signal-to-noise ratio of the data is low and the uncertain continuum level plays a role.
If the line is optically thin, there is a clear mapping from the model $\tau$ and $T$ to the observed line width and the ratio of narrow to broad line flux.
As noted above, \Ha\ frequently has a P-Cygni profile, which indicates the line formation region is not optically thin and affects the narrow line flux.
Therefore, our fitting procedure was focused on the broad component, and $\tau$ can only be loosely constrained by the physical requirements on the gas temperature.
For the level of ionization indicated in SNe IIn, a gas temperature $\sim 10,000 - 20,000$ K is expected \citep{kallman82}.
Our method depends on being able to separate the broad component from the narrow component of \Ha.  We found that by plotting the line flux on a logarithmic scale, it was generally possible to separate these components because of an inflection at the transition point in the line profiles.
The inflection did not stand out on a linear scale.  We plot the spectra here on a log scale.

The thickness of the medium $\eta$, expanding wind velocity $v_w$ and $v_{sh}$ determine the asymmetry of the broad component, the shift in the peak of the line and the shape of the narrow component.
The value of $v_w$ suggested by the P-Cygni profile, if available, was used in the fits.

In addition to the scattering calculation, a critical part of the fitting procedure is choosing the continuum level because it is crucial for determining the outer wings of the line.
When possible, our fits extend out to a wavelength region where the scattered line does not appear to be playing a role.
The observed spectra were corrected for the reddening listed in Table~\ref{params}, where the ratio of total to selective extinction $R_V=3.1$.
The continuum is fit by a blackbody spectrum.
In many cases, the full continuum cannot be represented by a single blackbody continuum.
In those cases, a blackbody that matches the continuum near the line was subtracted. 

Another parameter of the models is the redshift $z$.
Because of the effect of the expanding medium, the redshift of the supernova cannot simply be determined by the peak of the narrow component.
If the redshift of the supernova is available, e.g. SN 1998S, this redshift is chosen.
In most cases, only the redshifts of the host galaxy are available, which can only loosely constrain the redshift of the supernova.
Then $z$ was set by the broad component of the spectrum.

\begin{table*}
\begin{minipage}{\textwidth}  
\begin{center}
\caption{Supernova observational parameters }
\label{params}
\begin{tabular}{lccccccc}
  \hline\hline
  SN & Date of & Age & Redshift   & Resolution & $T_b$ &  $E(B-V)$ & Ref. \\
     & Observation &  (days) & $z$    &  (\AA)  &  (K) & &\\ \hline
  \multirow{2}{*}{1998S} & 1998 Mar 4 & 1.9  & \multirow{2}{*}{$0.00286$}  & 0.2 & 28,000 & \multirow{2}{*}{0.23 } & (1)\\ \cline{2-3}\cline{5-6}\cline{8-8}
                         & 1998 Mar 6 &  4   &   &  8  & 28,000 &  & (2),(3)\\ \hline
  2005cl &  2005 Jul 16 &  44 & $0.0259$ & 5 & 19,000  & 0.4   & (4) \\ \hline
  2005db &  2005 Aug 14 &  36 & $0.0151$ & 5 &  6200   & 0.3   & (4)  \\ \hline
  2012bq &  2012 Apr 12 &  13 & 0.0415   &18 & 12,000  & 0.2   &  (5)\\ \hline
  2005gj &  2005  Dec 2 &  71 & $0.0616$ & 3 & 10,000  & 0.4   & (6)  \\ \hline
  2008J  &  2008 Jan 17 &   2 & $0.0159$ & 7 & 10,000  & 0.8   & (7) \\ \hline
  2008cg &  2008 May 5  &   3 & $0.0362$ &11.6& 9000   & 0.2   &  (6) \\ \hline
  2009ip &  2012 Oct 14 &  21 & $0.00572$&1.3& 11,000  & 0.019 & (8)  \\ \hline
  2010jl &  2010 Nov 15 &  36 & 0.0107   &4.3&  5500   & 0.058 & (9)  \\ \hline
  2011ht &  2011 Nov 11 &  43 & $0.0036$  & 7& 13,000  & 0.062 & (10)  \\ \hline
\end{tabular} 
\end{center}
\emph{Note.} The ages are from the time of discovery.  The listed redshifts of SNe 2005cl, 2005db, 2005gj, 2008J, 2008cg, 2010jl, and 2011ht are the measured values for the host galaxies.  The redshifts of SNe 2012bq and 2009ip are measured from the peaks of the Balmer emission lines of the supernovae.  The spectrum of SN 2011ht is redshift corrected.

\emph{References.} (1)  \cite{shivvers14}; (2)  \cite{leonard00}; (3)  \cite{fassia01}; (4)  \cite{kiewe12};  (5) The spectrum is not published but is available in the WISeREP database; (6) \cite{silverman13}; (7)   \cite{taddia12};  (8)  \cite{margutti14}; (9)  \cite{borish13};  (10) \cite{humphreys12}
\end{minipage}
\end{table*}

\begin{table*}
\begin{center}
\caption{Supernova model parameters used in fits}
\label{fit_param}
\begin{tabular}{lccccccc}
  \hline\hline
  SN & Model  & FWHM & Radius  &$\tau$  & $T_e$  &  $v_w$ & $v_{sh}$ \\
     & redshift  &   (km s$^{-1}$)  &   ratio $\eta$ &  & (K)  & (km s$^{-1}$) &   (km s$^{-1}$) \\ \hline
  1998S (day 2) &  0.00286  &  1480 & 3  & 1.6 & 22,000  & 40 & 200 \\ \hline
  1998S (day 4) &  0.00286  &  2490 & 3  & 4   & 24,000  & 40 & 500 \\ \hline
  \multirow{2}{*}{2005cl}   &  0.0275 &  2930 & 2  & 1.5 & 80,000  & 800 & 0  \\ \cline{2-8}
      & 0.0293 & 2630 & 1.2 & 5  & 18,000 & 1300 & 0 \\ \hline
  \multirow{2}{*}{2005db} &  0.0163   & 1840 &  1.4  &  2 & 24,000  & 900 & 0 \\ \cline{2-8}
      &  0.0158   & 1700 &  1.4  &  6 & 6000  &  500  &  0 \\ \hline
  2012bq & 0.041   & 2650  &  2  & 6  & 13,000 & 1000  & 0  \\ \hline
  2005gj &  0.0621  &  1240  & 2  &  1.3 & 16,000  & 300 & 0 \\ \hline
  2008J  &  0.0162  &  1030  & 2  &  2.5 & 7000    & 200 & 0 \\ \hline
  2008cg &  0.036   &  1120  & 2  &   2  & 10,000  & 150 & 0 \\ \hline
  2009ip &  (0.00615)$^a$ & (1280) & (2)  & (1.6)  & (14,000)  & (250)  & (0) \\ \hline
  2010jl &  0.0108  & 1830   & 3 &  3  & 18,000  & 100 & 0 \\ \hline
  2011ht &  0.0041  & 2010  & 1.5  &  4.5 & 12,000  & 600 & 0 \\ \hline 
\end{tabular} 
\end{center}
\emph{Note.} $^a$ The fit parameters are in parentheses because there is not a reasonable electron scattering model fit in this case.
\end{table*}

\subsection{SN 1998S}

\begin{figure}
\includegraphics[width=\columnwidth]{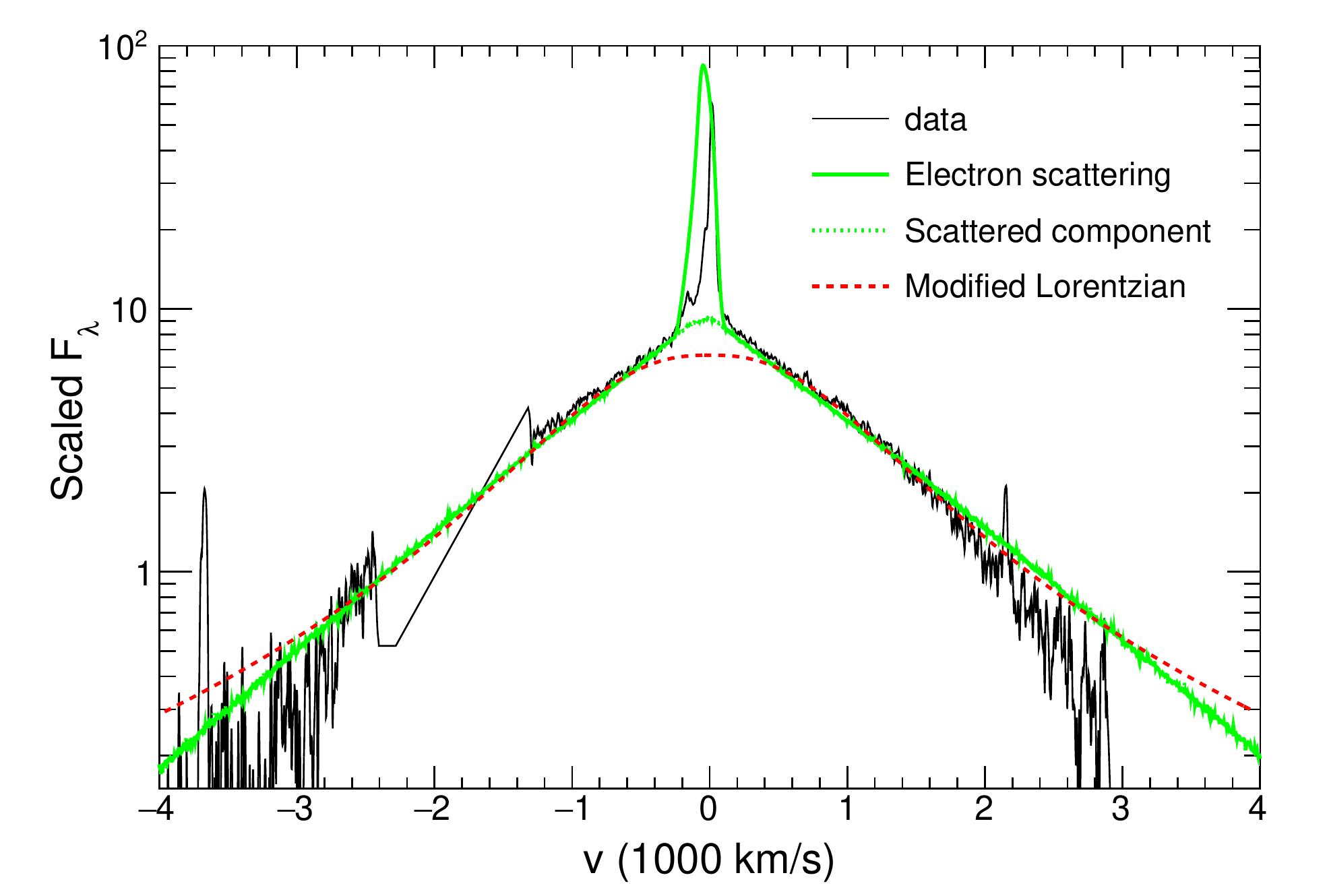}
\caption{ Comparison of the Monte Carlo electron scattering model result and a modified Lorentzian profile with the SN 1998S \Ha\ emission line on 1998 March 4 \citep{shivvers14}.  The narrow lines near $+2200\kms$ and $-3700 \kms$ are due to [\ion{N}{ii}] $\lambda$6611 and $\lambda$6482.  See the text and Table \ref{fit_param} for details.
}
\label{98s4}
\end{figure}

\begin{figure}
\includegraphics[width=\columnwidth]{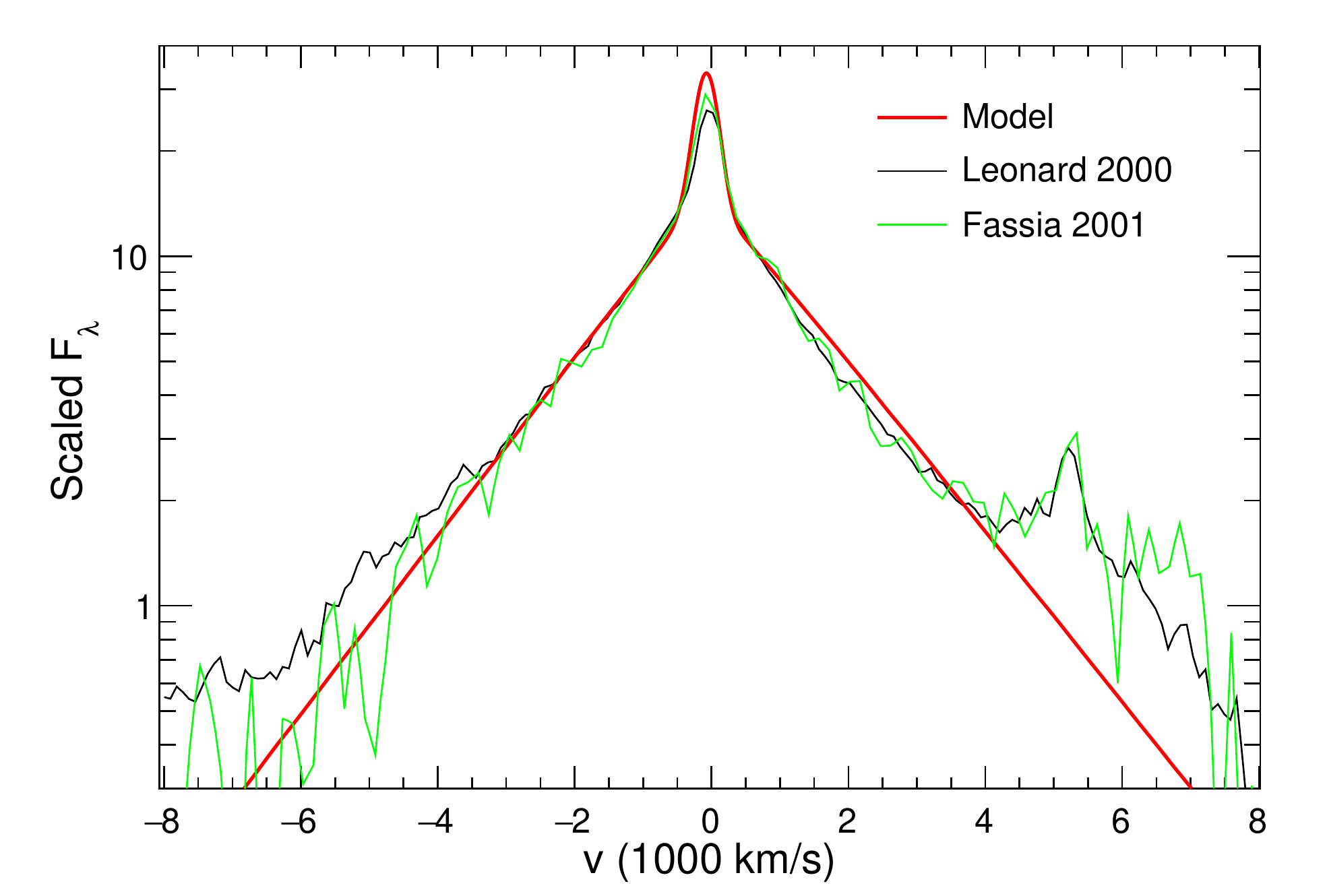}
\caption{ Comparison of the electron scattering model  result with the SN 1998S \Ha\ emission line on 1998 March 6 \citep{leonard00}.
The \ion{He}{i} $\lambda$6678 line is present, with a broad component. 
}
\label{98s}
\end{figure}

SN 1998S was discovered on 1998 March 2.7, which was probably
within a few days of the explosion \citep{fassia01}.
Spectra taken between March 4 and March 7 \citep{shivvers14,leonard00,fassia01} showed evidence for a narrow \Ha\ line with broad wings
(Figs.\ \ref{98s4} and \ref{98s}).
We present spectra from March 4 and 5 separately because there was significant evolution of the broad component over that time.
For Table~\ref{params}, we took the redshift $z$ at the position of the supernova from \cite{shivvers14} because of the high spectral resolution in their observation.
The reddening and blackbody temperature are from \cite{fassia01}.
The narrow component in the spectra implied a wind velocity of $40\kms$ \citep{fassia01,shivvers14}.

\cite{chugai01} developed an electron scattering model for the broad line component in SN 1998S.
He fit the \Ha\ line in the spectrum from 1998 March 6 \citep{fassia01}.
In his Model A, \cite{chugai01} took an electron temperature of $21,700$ K, as determined from
the shape of the optical continuum.  The corresponding scattering optical depth was 3.4.
The model has an outer to inner radius ratio of $\sim 2$ for the dense scattering region.   The limited extent
of the dense region is indicated by the light curve of the supernova.

\cite{shivvers14} present a high resolution (0.2 \AA) spectrum of SN 1998S for 1998 March 4.
Their fig.\ 7 shows the \Ha, H$\beta$ and H$\gamma$ lines, described by the sum of a narrow Gaussian and a broad modified Lorentzian, where the exponent is allowed to deviate from 2.0.
In Fig.~\ref{98s4}, we model the high resolution observation from 1998 March 4, within 2 days of discovery \citep{shivvers14};
note that there are some gaps in the echelle spectrum where a straight line joins the data points.
The model parameters are in Table~\ref{fit_param}.  
The electron scattering model gives a good fit to the observed spectrum.  
A modified Lorentzian profile with an exponent 2.5 is also shown.
The modified Lorentzian roughly agrees with the observed profile out to $2500~\kms$ in the line wing, but it  bends up further out the line wing, which is not seen in the observed spectrum, and there is a significant difference with the broad electron scattering case at $v=0$.
The observed line profile is symmetric, implying that enhanced red emission due to a wind is not present.
This can be attributed to 2 factors:
the wind velocity, $40\kms$, is much smaller than the thermal velocities of electrons and
the extent of the scattering region is relatively small.  

In Fig.~\ref{98s}, we show fits to the spectra on 1998 March 6~\citep{leonard00} and on 1998 March 6~\citep{fassia01}.
The latter one is that used by \cite{chugai01}, 
The signal-to-noise ratio is higher in the Keck LRIS spectrum with a spectral resolution of $\sim 8$ \AA\ \citep{leonard00}.
The FWHM of the Gaussian smooth function ($\sigma=110\kms$) is consistent with
the spectral resolution of the observation.

The figures show that the electron scattering model accounts well for the line profile, as deduced by \cite{chugai01}.
The profiles show the approximate exponential shape that is characteristic of electron scattering, as discussed in Section 2.
Most notable is the large increase in the FWHM of the broad \Ha\ component from day 2 to day 4, which requires an increase in scattering optical depth and/or an increase in electron temperature.
Rapid evolution is also present in the \ion{He}{i} $\lambda$6678 line: no broad component is present on March 4,
but it is present on March 6 (Figs.\ \ref{98s4} and \ref{98s}).
\cite{shivvers14} note that the \ion{He}{i} lines show no trace of broad wings on March 4 although they are as strong as the \ion{He}{ii} line,
which does show a broad feature.

Fig.\  \ref{98s} shows that the blue wing is somewhat higher in the \cite{leonard00} spectrum than in the \cite{fassia01} spectrum, although the spectra are close in time.
A possibility is that the line profile showed rapid evolution.
We found that the excess blue emission could not be fitted in the context of an electron scattering model.
As a possible source of the emission, we note that some \Ha\ photons from the \Ha\ shock
region of the supernova may be able to escape through the circumstellar envelope.  
At later times, there is shock emission extending to the blue by $\ga 4,000\kms$ \citep{fassia01}.

\subsection{SN 2005cl, SN 2005db, and SN 2012bq}

\begin{figure}
\includegraphics[width=\columnwidth]{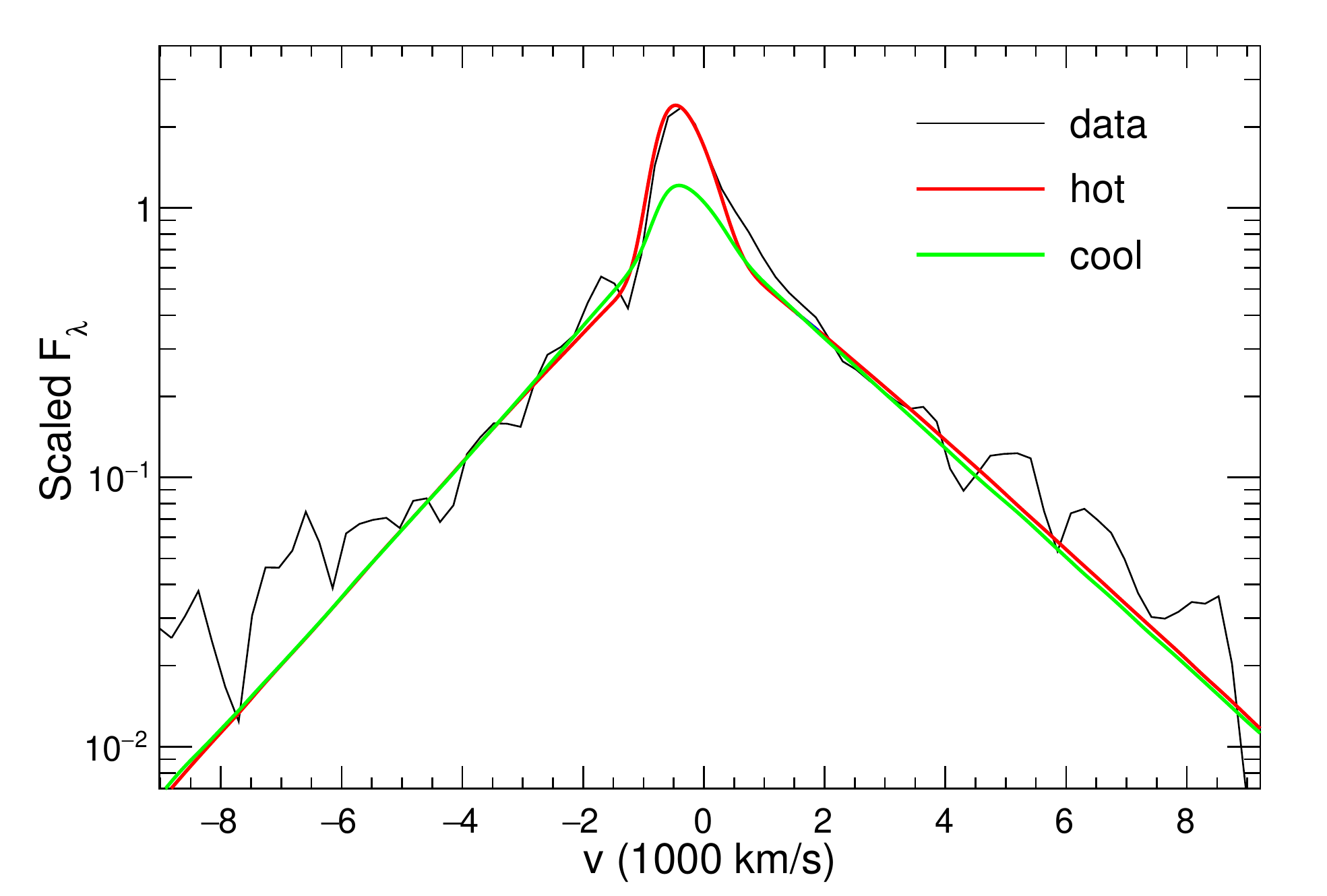}
\caption{ Comparison of the electron scattering model result with the SN 2005cl \Ha\ emission line on 2005 July 16 \citep{kiewe12}.
}
\label{05cl2n}
\end{figure}

These events have some properties in common: high FWHM, high wind velocity from P Cygni line, and relatively high redshift.
The Type IIn supernovae SN 2005cl and SN 2005db were observed as part of the Caltech Core Collapse supernova Project (CCCP), which
followed up on every core collapse supernova observable from Palomar Observatory during the time of the project \citep{kiewe12}.
The observed events may thus be typical of Type IIn supernovae.
We include SN 2012bq in this group because it has similar properties.

A spectrum of SN 2005cl obtained on 2005 July 16,  44 days after discovery 
\citep{kiewe12}, is shown in Fig.\ \ref{05cl2n}.
The spectrum shows a P Cygni profile in the narrow \Ha\ line, which \cite{kiewe12}
take to indicate an unshocked wind velocity of $1318\pm 223\kms$.
Two models for SN 2005cl were developed and are shown in Fig.\ \ref{05cl2n}.
The first one was intended to fit the whole spactrum, including the narrow \Ha\ component.
An acceptable fit is obtained but the electron temperature of 80,000 K is unreasonably high, so
we also give a fit with a cool electron temperature.
In that case, the narrow component is not well fit, but our model is not intended to fit the narrow component.
Also, the electron scattering optical depth is high, $\sim 5$.

\begin{figure}
\includegraphics[width=\columnwidth]{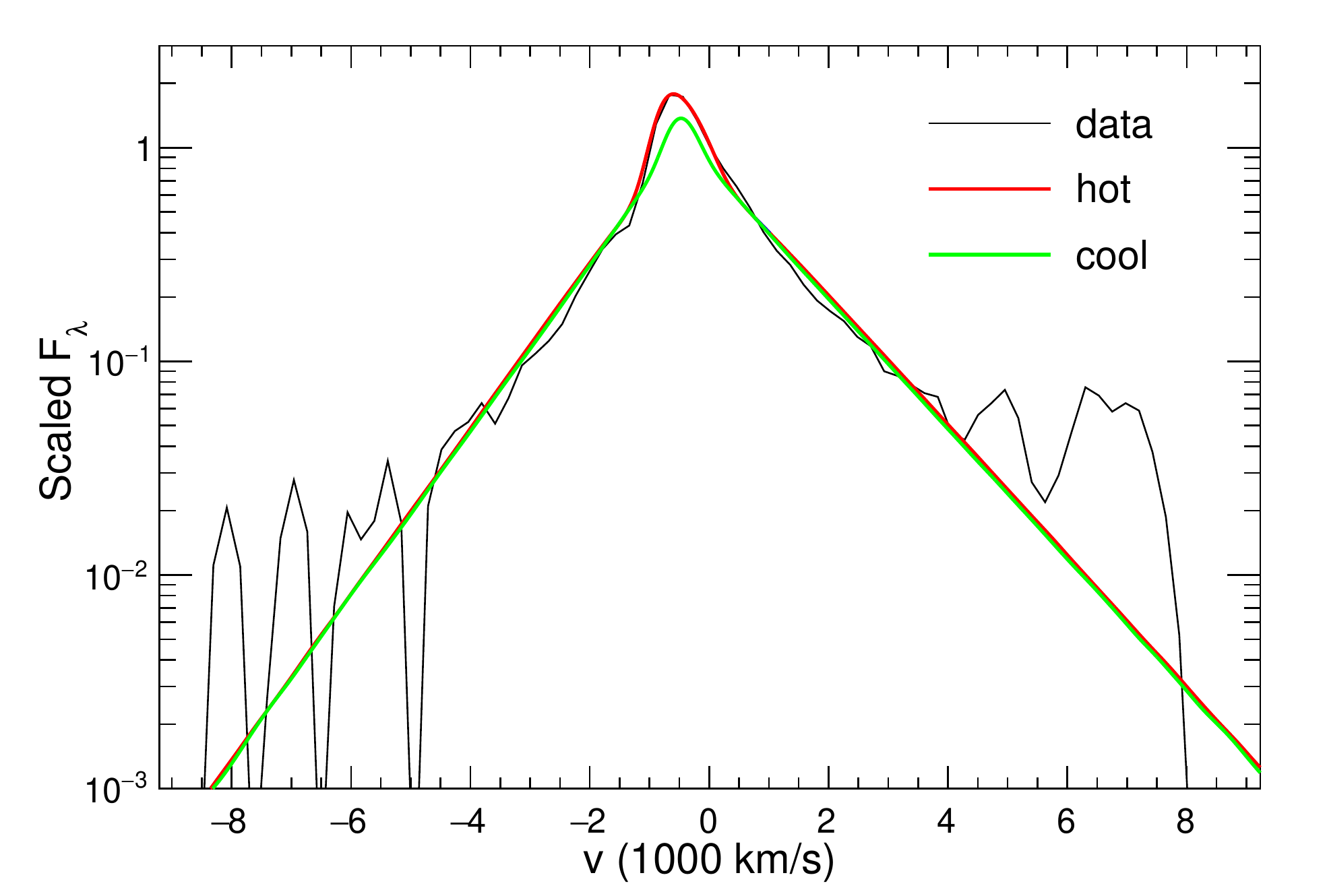}
\caption{Comparison of the electron scattering model result with the SN 2005db \Ha\ emission line on 2005  August 14 \citep{kiewe12}.
The parameters for the two models are given in Table \ref{fit_param}.
}
\label{05db2n}
\end{figure}

A spectrum of SN 2005db was obtained on 2005 Aug 14,  36 days after discovery      
\citep{kiewe12}, see Fig.\ \ref{05db2n}.
The spectral resolution was $\sim$ 5 \AA.
The spectrum shows a P Cygni profile in the \Ha\ line, which \cite{kiewe12}
take to indicate an unshocked wind velocity of $1113\pm 65\kms$.
For SN 2005db, the observed profile does not show a clear turning point at the joint of narrow component and broad component,  which introduces
some uncertainty into our modeling. 

\begin{figure}
\includegraphics[width=\columnwidth]{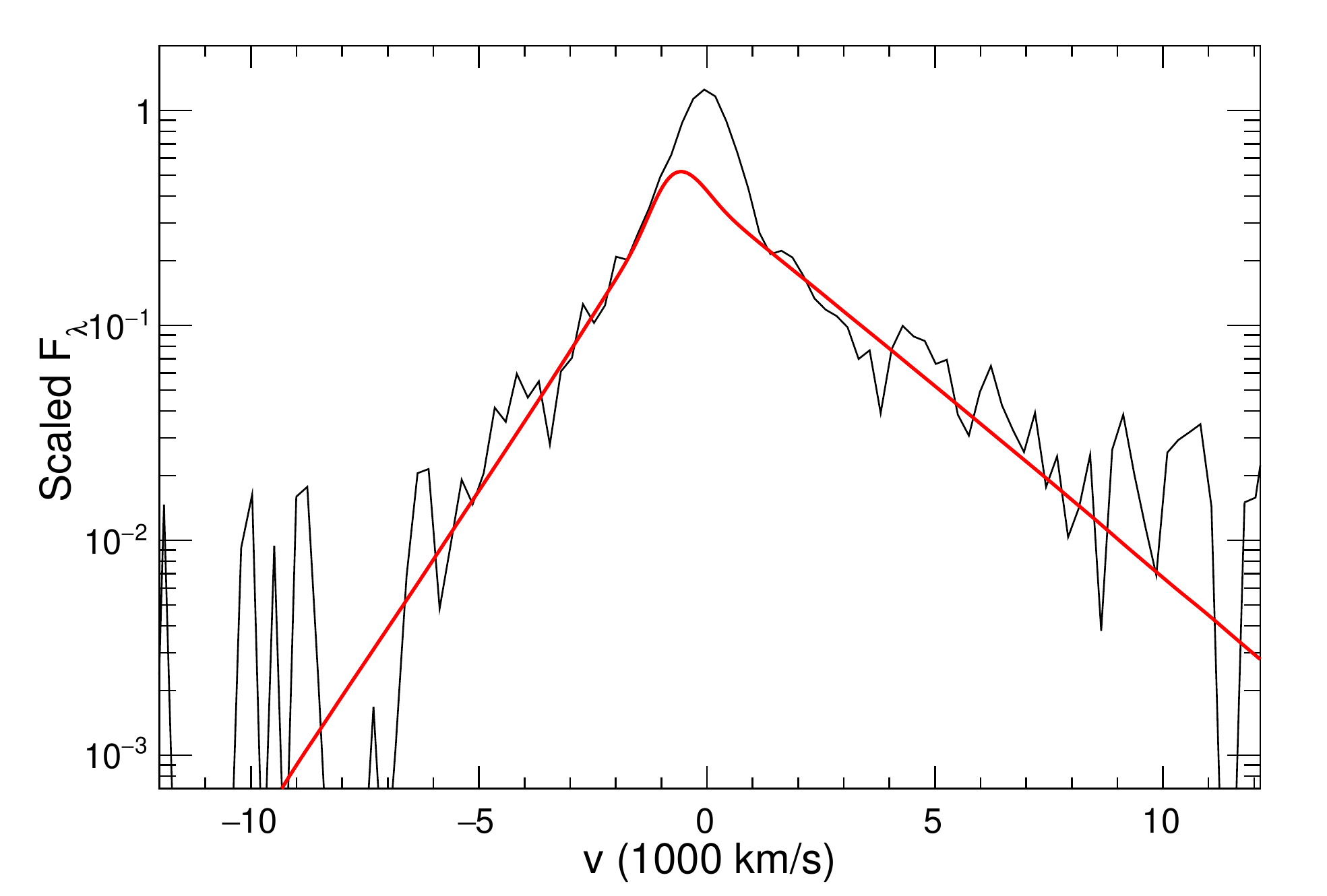}
\caption{ Comparison of the electron scattering model result with the SN 2012bq \Ha\ emission line on 2012 April 12  (see text for details).
}
\label{12bq}
\end{figure}

SN 2012bq was discovered on 2012 March 30 \citep{drake12}.
It was studied as part of the PESSTO project SSDR1 (Spectroscopic Survey Data Release 1) and is listed
 in WISeREP  also as  LSQ12bry.  
The observation in  Fig.\ \ref{12bq}  is from 2012 Apr 11
with the NNT 3.58 m telescope, $\sim 18$ \AA\ resolution.

 The situation is similar for these 3 supernovae.
 To have a reasonable electron temperature, the narrow component is poorly fit and the value of $\tau \sim 5-6$.
 In all 3 cases, there is some asymmetry of the broad component, with stronger emission on the red side.
 As discussed in Section~\ref{sec:wind},  the asymmetry can be due to scattering in the outflowing wind.
 In all 3 supernovae there is evidence for a wind with speed $\sim 1000\kms$, which is consistent with what is needed to fit the observations.
 While there is asymmetry, the line shape of the broad component is still exponential.
 
 A possible problem with this scenario is that of the 4 Type IIn supernovae that \cite{kiewe12} observed, one of them, SN 2005cp, had    
 emission that skewed the line profile to the blue, inconsistent with scattering in an outflowing wind.
 However, the emission is to the blue and the line profile is not consistent with an exponential.
 In this case, we conjecture that blue emission is due to the systematic Doppler shift of the emission. 
We did not model spectra of the fourth Type IIn supernova observed by \cite{kiewe12}, SN 2005bx, because of the poor signal-to-noise.

\subsection{SN 2005gj, SN 2008J, and SN 2008cg}

The supernovae SNe 2005gj, 2008J, and 2008cg belong to the class of Type Ia-CSM supernovae, which
show clear features of a Type Ia spectrum, as well as Type IIn characteristics \citep{hamuy03,silverman13}.
The Type Ia features typically grow stronger with age of the supernova.

\begin{figure}
\includegraphics[width=\columnwidth]{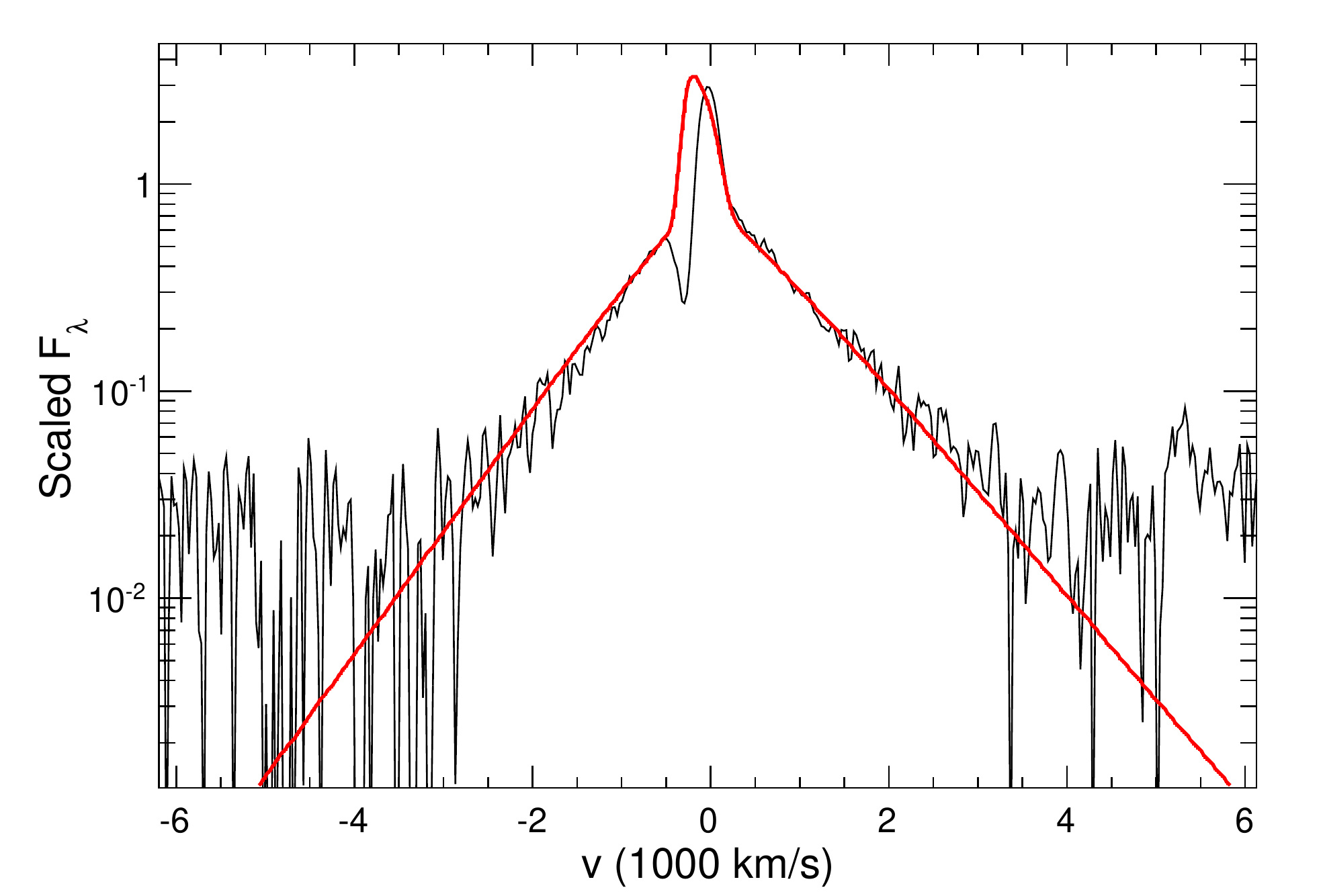}
\caption{Comparison of the electron scattering model result with the  SN 2005gj \Ha\ emission line on 2005 December 2 \citep{silverman13}.
}
\label{05gjn}
\end{figure}

SN 2005gj was discovered on 2005 September 26 when it had an estimated age of 4 days since explosion \citep{aldering06}.
\cite{aldering06} show fits of an electron scattering model to spectra on days 11, 64, and 71 after the explosion date;
the fits are reasonable.
Fig.\ \ref{05gjn} shows a spectrum taken with Keck II on 2005 December 2 with a resolution of $\sim 3$ \AA\
\citep{silverman13};  the age is 67 days after discovery, or 71 days after explosion.
The spectrum at that time shows a very broad feature roughly centred on \Ha\ that cannot be explained with
the electron scattering model.
Fig.\  7 of \cite{aldering06} makes clear that the feature is connected with a SN 1991T-type spectrum, a luminous subclass
of Type Ia supernovae.
Fig.\ \ref{05gjn} shows a comparison with our Monte Carlo electron scattering model.
In addition to the blackbody spectrum, a Gaussian model of the Type Ia feature near the \Ha\ line has been subtracted off.  
The Gaussian subtraction parameters applied are in Table \ref{Gaussian_param}.
The wind velocity from the P Cygni feature is estimated to be $\sim 300 \kms$ \citep{aldering06,prieto07}, and we used that value in our model.
There is a small asymmetry in the observed broad line profile that is captured by the model with a wind velocity.

\begin{table}
  \centering
  \caption{Gaussian model parameters used to subtract the Type Ia feature near the \Ha\ line}
  \label{Gaussian_param}
  \begin{tabular}{lccc}
    \hline\hline
    SN& Mean (\AA) & Width (\AA) & Amplitude \\ \hline
    2005gj & 6540 & 100 & 0.34 \\
    2008J  & 6527 & 125 & 0.25 \\
    2008cg & 6540 & 125 & 0.06 \\ \hline
  \end{tabular}
\end{table}

\begin{figure}
\includegraphics[width=\columnwidth]{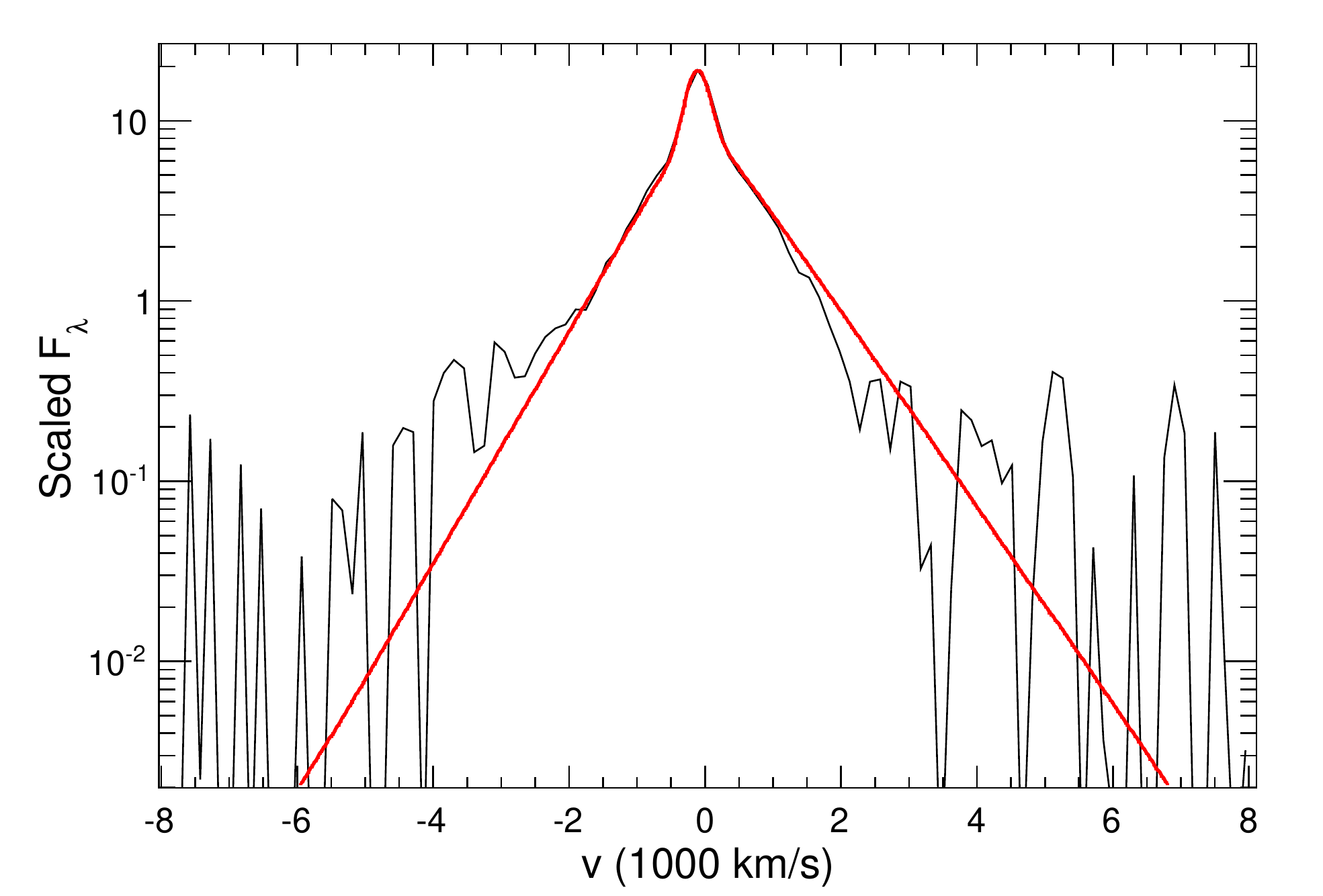}
\caption{Comparison of the electron scattering model result with the  SN 2008J \Ha\ emission line on 2008 January 17   \citep{taddia12}.
}
\label{08jn}
\end{figure}

The spectrum of SN 2008J in Fig.\ \ref{08jn} is from 2008 January 17, which is 2 days after the discovery of the supernova and 5.8 days before maximum $B$ light \citep{taddia12}.  The spectral resolution is $\sim 6-9$ \AA.
\cite{taddia12} find that the SN Ia features that appear are of the SN 1991T - type, as in SN 2005gj.
Any asymmetry in the line is small and our model has a small wind velocity.

\begin{figure}
\includegraphics[width=\columnwidth]{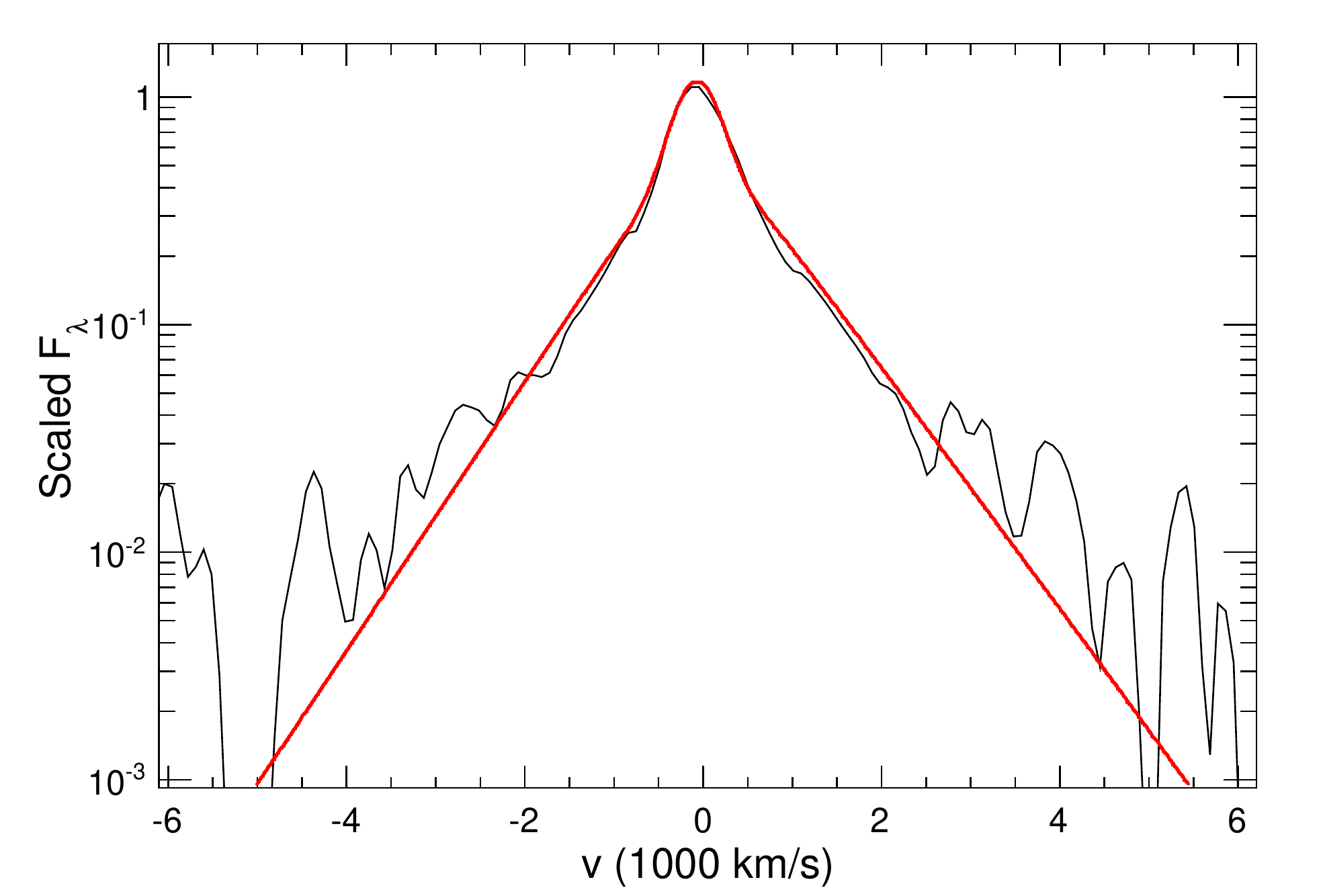}
\caption{Comparison of the electron scattering model result with the  SN 2008cg \Ha\ emission line on 2008 May 5   \citep{silverman13}.
}
\label{08cgn}
\end{figure}

Fig.\ \ref{08cgn} shows a spectrum of SN 2008cg taken at the Lick 3m telescope on 2008 May 8,
which \cite{silverman13} estimate to be 9 days after maximum light, or 3 days from discovery.
The supernova is identified by \cite{silverman13} as Type Ia-CSM.
This is not so clear from the early spectrum, but at later times the evidence seems quite clear.
\cite{silverman13} note that initially the spectrum of SN 2008cg resembles a normal Type IIn event
but after about 2 months, it closely resembles SN 2005gj.

We thus find that SNe Ia-CSM have \Ha\ line wings that can be fit by an electron scattering profile.
The line profiles are either symmetric or show a small asymmetry to the red, and are superposed on broad Type Ia supernova features that grow in
strength with age.
\cite{silverman13} find that the \Ha\ lines in most SNe Ia-CSM show a deficit in the red wing
starting at $\sim 75-100$ days after maximum light, later than the ages of the spectra considered here.
This development is commonly attributed to the formation of dust which absorbs emission from gas moving away
on the far side of the supernova.
This interpretation is not compatible with line formation by electron scattering by thermal electrons, so the 
supernovae would have to begin a new phase of evolution.

\subsection{SN 2009ip}

\begin{figure}
\includegraphics[width=\columnwidth]{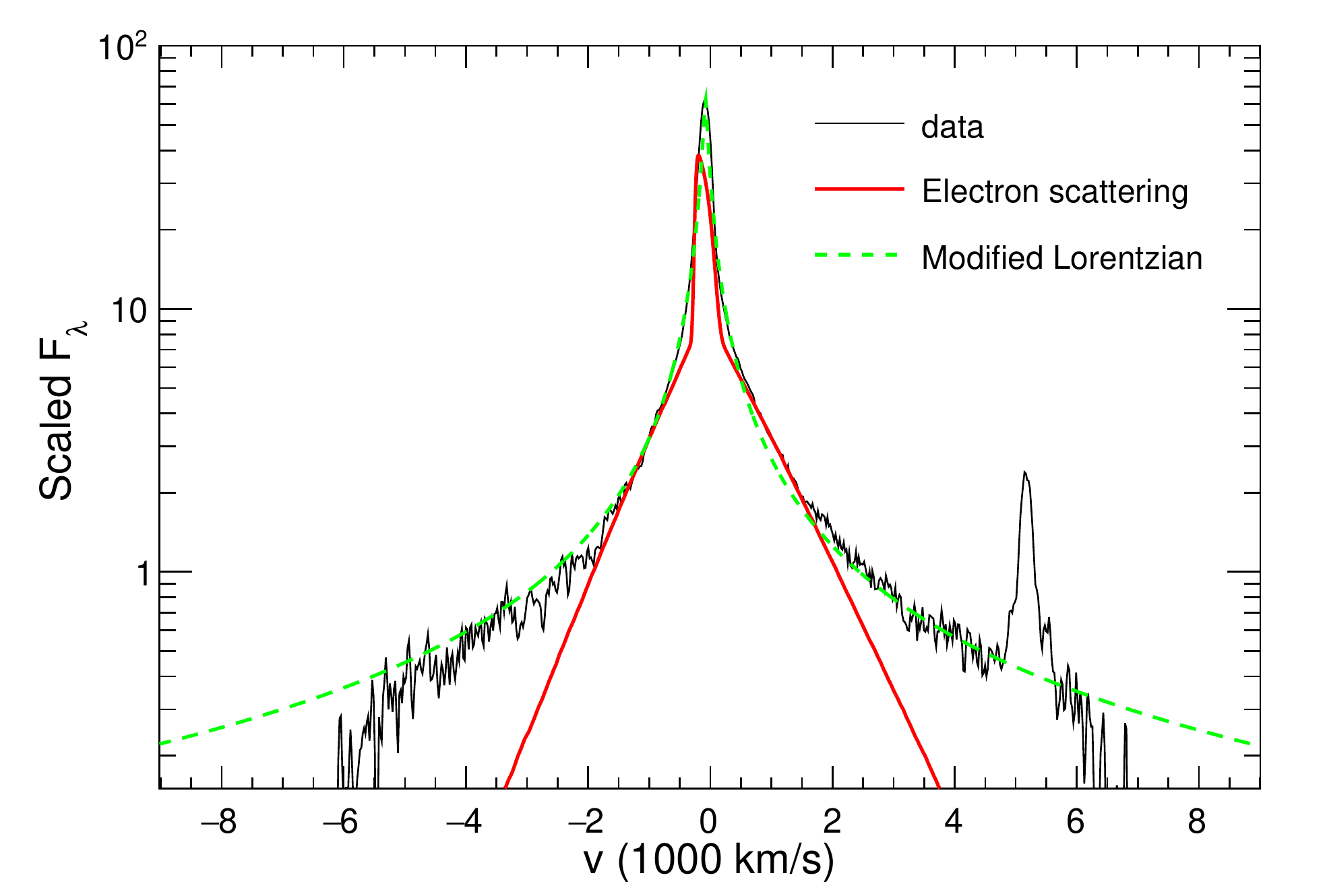}
\caption{Comparison of the electron scattering model result and the modified Lorentzian with the SN 2009ip \Ha\ emission line on 2009 October 14~\citep{margutti14}.
The electron scattering model does not provide a good fit.
}
\label{09ip}
\end{figure}

SN 2009ip is a massive star that underwent an unusual period of bursting activity covering years~\citep{mauerhan13,margutti14}.
There was an outburst in 2012 September, known as the 2012b event, in which it reached an absolute magnitude of $-18$ and showed emission line velocities of $\sim 10^4\kms$, which are characteristic of supernovae.
\cite{smith14} find that persistent broad emission lines in the spectrum require an ejecta mass and kinetic energy that are characteristic of supernovae.
However, \cite{fraser15} find no conclusive evidence for a core collapse supernova from the time of the outburst to 820 days later.
Fig.\  \ref{09ip} is from the Multiple Mirror Telescope (MMT), with spectral resolution $R=5000$ on 2012 Oct 14~\citep{margutti14}.

The emergence of the  \Ha\ line was presumably related to shock interactions in the circumstellar medium.
We were unable to fit the wings on the \Ha\ line with an electron scattering model.
The excess in the line wing cannot be explained by an uncertainty in the blackbody background. 
The line wings do not have the approximately exponential profile expected for electron scattering.
In addition, there is no inflection observed between the line core and line wing as expected in the electron scattering model, even though the narrow component is well resolved.
However, the entire line profile is close to a modified Lorentzian profile although there is no physical rationale for this profile.
A modified Lorentzian profile with a power law index 1.2 is shown in green dashed line in Fig.\  \ref{09ip}.

\subsection{SN 2010jl}

\begin{figure}
\includegraphics[width=\columnwidth]{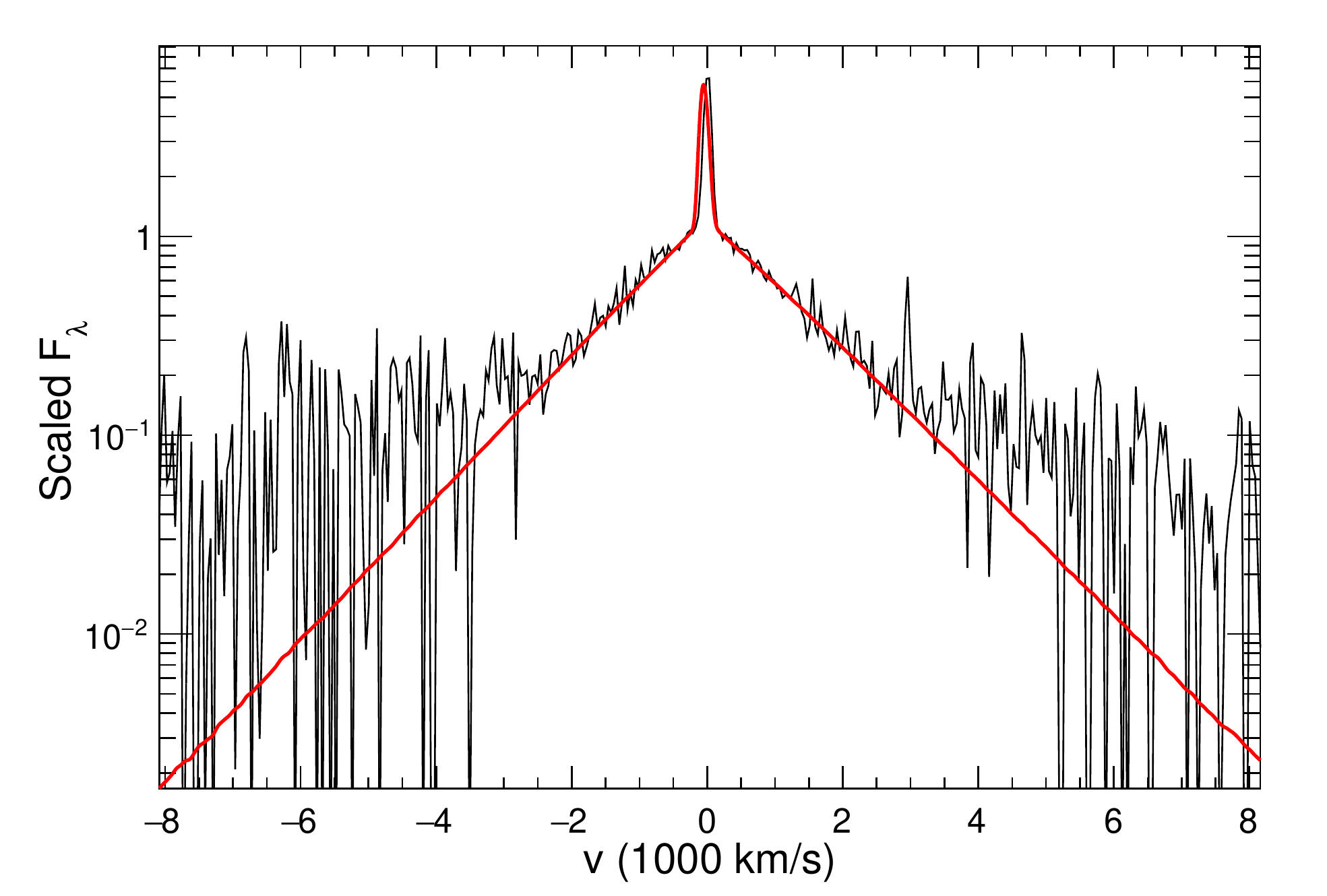}
\caption{Comparison of the electron scattering model result with the  SN 2010jl \Pb\ emission line on 2010 November 15 \citep{borish13}.
}
\label{10jl}
\end{figure}

SN 2010jl was an especially well observed Type IIn supernova, showing line profiles that are in agreement with electron scattering \citep{fransson13,borish13}.
The \Pb\ line has a smoother background continuum than the \Ha\ line, and the P-Cygni feature present in \Ha\ does not show up in the narrow component of \Pb.  We thus chose the \Pb\ line to model in this case.  

Fig.\ \ref{10jl} shows the \Pb\ line.
The narrow \Ha\ line observed in SN 2010jl indicates $v_w=100\kms$ \citep{fransson13}.
This small wind velocity is consistent with the observed approximately symmetric scattering wings.
\citet{fransson13} and \citet{borish13} show that the broad components of Balmer lines shift to the blue, but the narrow lines do not shift in the later time spectrum.
It is plausible that the scattering region is distinct from the narrow line production.
The evolution suggests that mass motions come to play a role in the line formation \citep[e.g.,][]{dessart15}.

 \cite{smith10} found a good fit to the early \Ha\ line with a modified Lorentzian, and suggested a moderate optical depth
to electron scattering.
We  have shown that an exponential is a better fit to electron scattering wings than a modified Lorentzian; the Lorentzian has a curvature
that is not expected for electron scattering.
We suggest that these disparate fits to the SN 2010jl spectrum are due to the uncertainty in the background subtraction for the \Ha\ line.
The background emission shows complex structure near the \Ha\ line.

SN 2010jl has been the target of multiwavelength observations that can be used to estimate the preshock column density of H.
The absorbing column for X-rays and the bolometric luminosity imply a column $N_H \sim (1-4)\times 10^{24}$ cm$^{-2}$ on 15 Nov 2010 \citep[Fig.\ 11 of][]{chandra15},
which corresponds to an electron scattering optical depth $\tau \sim 0.7-2.7$.
The value of $\tau$ needed for the line profile is marginally higher, which could indicate asymmetry in the emission region \citep{fransson13}.

\subsection{SN 2011ht}

\begin{figure}
\includegraphics[width=\columnwidth]{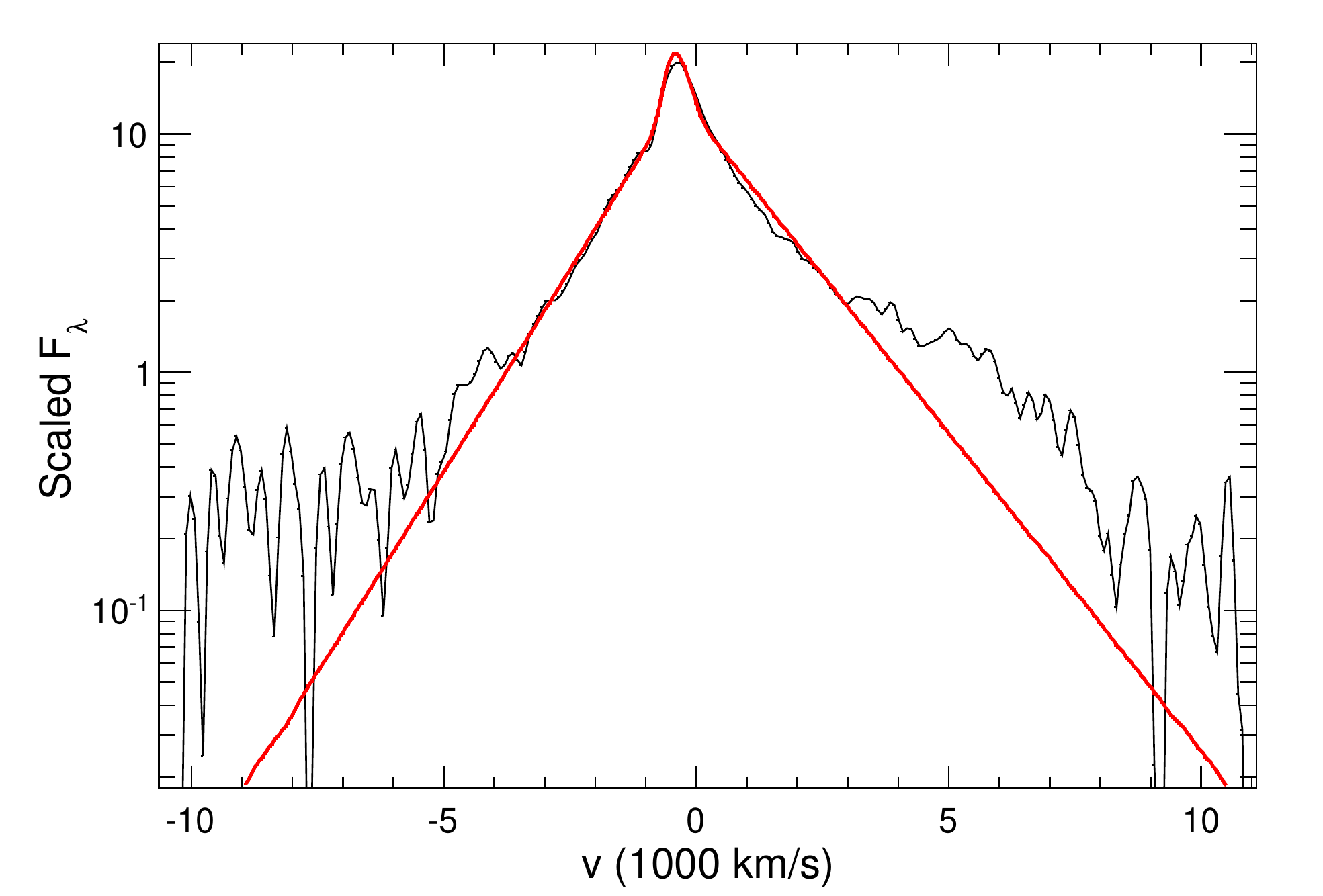}
\caption{ Comparison of the electron scattering model result with the SN 2011ht \Ha\ emission line on 2011 November 11  \citep{roming12}.
}
\label{11ht}
\end{figure}

We modeled an \Ha\ spectrum of SN 2011ht obtained on 2011 Nov 11 from the Apache Point Observatory (APO) 3.5 m \citep{roming12}, which was the earliest spectrum available. 
Discovery was on 2011 Sept 29, so this is moderately late.  
\citet{humphreys12} suggested that this may not be a supernova, but rather a giant eruption event, because the kinetic energy of its ejecta seems much less than a regular supernova.

The Balmer emission lines show very broad wings with some asymmetry to the red, which is a classic electron scattering signature.  
Later higher resolution spectra showed clear evidence for P Cygni line and a wind velocity of $600\kms$ \citep{humphreys12}.
Fig.\ \ref{11ht} shows a fit with $v_w=600\kms$ that is in good agreement with the observed asymmetry.
The spectrum from WISeREP has been corrected for the galactic redshift of 0.0036;
an extra redshift of 0.0005 was applied in the fitting.
The SN 2011ht spectrum has two strong and very broad \ion{He}{i} emission lines at 5876 \AA\ and 7065 \AA\  \citep{humphreys12},
suggesting that the excess on the red wing from $3000\kms$ to $7000\kms$ is likely to be a broad \ion{He}{i} 6678 feature.

\cite{chugai16} suggested that the situation in SN 2011ht may be  different from the standard model described here, based on the close correspondence between SN 2011ht and SN 1994W.
Line profiles in SN 1994W were initially modeled with 
electron scattering in an outflowing circumstellar medium \citep{chugai04}.
However, \cite{dessart09} noted that there was no sign of high velocity gas at later times, which would be expected if there were a normal energy
supernova inside the circumstellar matter.
 \cite{dessart09} identified the velocity of the P Cygni feature, $\sim 800\kms$, with the velocity of the photosphere and not with circumstellar gas.
 In this model, electron scattering in the photospheric region gives rise to the broad wings on the H$\alpha$ line.
 In addition to the lack of high velocities, this model can better explain the strong wings on the H$\gamma$ line.
 \cite{chugai16} noted that the same arguments applied to the case of SN 2011ht.
Our work shows that the roughly exponential line profile wings occur widely when electron scattering is operating, so the presence of
the line wings may not require a particular model.

\section{DISCUSSION AND CONCLUSIONS}

The comparison of the electron scattering model with observations shows that scattering provides a
plausible explanation for the line profile wings in many supernovae designated as Type IIn.
It has been noted on occasion that a Lorentzian or modified Lorentzian profile gives a good approximation to
the profiles observed in SNe IIn \citep{leonard00,smith10,shivvers14}, but there is no physical reason to expect such a profile.
There has also been fitting by multiple Gaussians \citep[e.g.,][]{kiewe12}, but again there is not a clear
physical explanation for such a profile.

An expectation of the electron scattering model is that there should be enhanced emission on the red side of
the line if the scattering occurs in an extended surrounding medium with an outflow velocity.
This feature is generally not observed.
In the case of SN 1998S, there is evidence for dense mass loss occurring a short time before the explosion
\citep{chugai01} and, thus, a limited extent scattering region which is consistent with the symmetry of the \Ha\ line.
The lines observed in SNe Ia-CSM are also symmetric over the first 2 months but, in this case, the supernova light
curves do not indicate a late mass loss phase.
The observations imply that the scattering remains in a fairly narrow region, although it is expanding outward with time.
In the case of the SN 2005cl group of objects, there is an asymmetry with stronger emission to the red and these
cases can be modeled with an outflowing scattering region.
The observation of the asymmetry may be related to the high circumstellar outflow velocities found in these objects
through their P Cygni profiles \citep{kiewe12}.

To a first approximation, the line profile resulting from electron scattering has an exponential shape.
At low optical depths, the line profile has a concave shape, while at high optical depth it is convex.
However, the differences only become clear far out in the wings, where it is not possible to obtain accurate 
observational data.
The observed line profile shapes are not able to determine the value of $\tau$, so in the models there is a 
degeneracy in $\tau$ and $T$ in fitting the observed line width.
If the line formation were optically thin in the line, the ratio of narrow line component to broad would break the degeneracy;
however, the common observation of P Cygni features in the narrow line and the narrow line shape show
that is not the case.
In the models presented here, we made the assumption that the narrow to broad line ratio equals the optically
thin ratio.
Most of the model fits have $T$ in the $5000 - 20,000$ K range, which is close to the temperatures expected
in the photoionized gas and implies that the narrow to broad line ratio is not far from the optically thin case.
However, models of the supernovae SN 2005cl, SN 2005db, and SN 2012bq yield a high temperature and relatively
low optical depth.
The implication is that the narrow to broad ratio is higher than it would be in the optically thin limit.

The determination of the outer wings of the lines depends sensitively on the assumed continuum.
We have found that it is necessary to cover a broad wavelength range to obtain a reliable continuum fit.
In the case of the SNe Ia-CSM, there is structure in the continuum due to the underlying Type Ia spectrum that
leads to uncertainty in the continuum fit.
The strength of the feature near \Ha\ grows with age.

\section*{Acknowledgements}
We thank  Phil Arras for supplying the basic Monte Carlo code and  for discussions, and
Nikolai Chugai for a helpful referee's report.
We made use of the WISeREP data repository.
The simulations in this work were carried out on the Rivanna computer cluster at the University of Virginia.
This research was supported in part by  NASA grants  NNX10AH29G, NNX12AF90G, NNX14AE16G, and NNX15AE05G.


\end{CJK*}
\end{document}